\documentclass[11pt]{article}
\usepackage{amsmath,amsthm,amsfonts,amssymb,amscd,bbm,bm,xcolor,graphicx,booktabs,subcaption,placeins}
\usepackage[utf8]{inputenc}
\usepackage[top=1.0in, bottom=1.0in, left=1.0in, right=1.0in]{geometry}
\usepackage[skip=10pt, indent=0pt]{parskip}
\usepackage[round]{natbib}
\usepackage[colorlinks=true, linkcolor=black, citecolor=blue, urlcolor=blue]{hyperref}
\usepackage{xurl} 
\hypersetup{breaklinks=true}

\usepackage{mathtools}
\mathtoolsset{showonlyrefs}

\title{A Survival Framework for Estimating Child Mortality Rates using Multiple Data Types}
\author{Katherine R. Paulson$^1$, Taylor Okonek$^2$, Jon Wakefield$^{1,3}$\vspace{1em}\\
\small$^1$Department of Biostatistics, University of Washington\\
\small$^2$Department of Mathematics, Statistics, and Computer Science, Macalester College\\
\small$^3$Department of Statistics, University of Washington}
\date{\today}

\begin{document}

\maketitle

\section*{Abstract}

Child mortality is an important population health indicator. However, many countries lack high-quality vital registration to measure child mortality rates precisely and reliably over time. Research endeavors such as those by the United Nations Inter-agency Group for Child Mortality Estimation (UN IGME) and the Global Burden of Disease (GBD) study leverage statistical models and available data to estimate child survival summaries including neonatal, infant, and under-five mortality rates. UN IGME fits separate models for each age group and the GBD uses a multi-step modeling process. We propose a Bayesian survival framework to estimate temporal trends in the probability of survival as a function of age, up to the fifth birthday, with a single model. Our framework integrates all data types that are used by UN IGME: household surveys, vital registration, and other pre-processed mortality rates. We demonstrate that our framework is applicable to any country using log-logistic and piecewise-exponential survival functions, and discuss findings for four example countries with diverse data profiles: Kenya, Brazil, Estonia, and Syrian Arab Republic. Our model produces estimates of the three survival summaries that are in broad agreement with both the data and the UN IGME estimates, but in addition gives the complete survival curve.

\section{Introduction}\label{sec:intro}

In 2023, approximately 4.8 million children died before their fifth birthday, and many of these child deaths are preventable with well-established interventions and health systems strengthening \citep{igme-2024}. The United Nations (UN) Sustainable Development Goals (SDGs) include targets for global reduction in child mortality rates by 2030, specifically, fewer than 25 under-five deaths per 1000 live births and fewer than 12 neonatal deaths per 1000 live births. However, many countries lack high-quality vital registration (VR) systems to measure child mortality rates with high precision. To monitor child mortality rates and track progress towards the SDG targets, research groups such as the United Nations Inter-agency Group for Child Mortality Estimation (UN IGME) and the Global Burden of Disease (GBD) study which is carried out by the Institute for Health Metrics and Evaluation (IHME) use statistical models to estimate annual child mortality rates, including neonatal, infant, and under-five mortality rates \citep{igme-2024,schumacher_global_2025}.

UN IGME uses separate models for each age group of interest. First, they use the Bayesian B-spline Bias-reduction (or ``B3") model originally published in \citet{alkema2014global} to estimate the under-five mortality rate (U5MR) in each country in the world. Then, the neonatal mortality rate (NMR) is estimated using a separate model that leverages a posited relationship between NMR and U5MR \citep{alexander2018global}. Finally, the infant mortality rate (IMR) is estimated using complete VR data when available, and using the log-quad model otherwise \citep{guillot_modeling_2022,explanatory-igme}. The log-quad approach is used instead of the data directly because data on infant deaths are often subject to age-heaping and are hence considered unreliable.

Before GBD 2023, the GBD used separate models for U5MR and for the probability of death in a set of narrower age groups, and then uses a scaling process to enforce internal consistency such that the component age groups aggregate correctly to the estimated U5MR \citep{schumacher_global_2024}. In the most recent iteration of the GBD (GBD 2023) a new model was adopted, which uses one framework to estimate all age groups including child and adult ages \citep{schumacher_global_2025}. However, this new approach still involves many separate computational steps, including a global likelihood-based step involving covariates, smoothing of residuals in space and time, and a post-hoc method for calibrating uncertainty intervals. This sequence is repeated twice to handle data at varying levels of granularity with respect to age.

The aim of this paper is to propose a novel survival framework to estimate the probability of survival for each year in a period of interest (e.g.,~1950--present). We combine a survival model with a Bayesian inferential approach. This strategy offers a number of advantages:
\begin{itemize}
    \item The rich array of survival models can be exploited -- we model survival times as continuous random variables.
    \item Within the Bayesian framework, likelihood terms can be specified to synthesize each data type used for national child mortality estimation within a single comprehensive model:~household surveys, VR, and other pre-processed estimates of mortality rates. 
    \item By using a survival model we enforce proper relationships between age groups -- the probability of death between birth and age $a$ is constrained to be monotonically non-decreasing with respect to $a$.
    \item Further,  we can constrain mortality risk to be non-increasing over age -- something demographers expect to be true before age five.
    \item A survival model can be used to predict probability of death for any age group of interest, making it easy to study age groups beyond those traditionally estimated. In addition to age-specific probabilities of death, many other summaries of mortality can be extracted from estimated survival functions. For example, we can study other parameters demographers typically include in life tables, various conditional probabilities, and more.
    \item By adopting an explicit survival modeling framework, it is straightforward to consider data complications such as interval censoring, truncation, age-heaping, and complex survey design, all with valid and clear statistical inference. 
    \item The standard inferential approach means that existing computational tools can be used.
\end{itemize}
Neither GBD nor the log-quad model use survival models or follow established and validated methods for statistical inference. Consequently, it is not possible to theoretically assess the statistical properties of these methods, including uncertainty interpretation.

Our work builds on previous literature either modeling child mortality with continuous age or explicitly using techniques from survival analysis to study child mortality. The log-quad model is one method which estimates continuous age schedules of child mortality, and it was used by UN IGME in 2024 to estimate the IMR in countries without high-quality VR \citep{guillot_modeling_2022,verhulst_divergent_2022,explanatory-igme,verhulst2025}. The log-quad model is a two-parameter model that borrows from model life table literature by leveraging the Human Mortality Database \citep{barbieri_data_2015} VR data to characterize commonly observed patterns of mortality. While the log-quad model is a continuous model, it is not translatable to a true survival model (meaning that it is possible to obtain implied survival functions that are not monotonically non-increasing). It is a curve-fitting method that is neither based on any likelihood construction nor developed non-parametrically to estimate a survival function. The log-quad model therefore does not benefit from either the strengths of traditional survival analysis or established approaches to statistical inference.
One consequence of this, for example, is that there is no evidence that the log-quad interval estimates have the correct coverage. It is not uncommon for public health researchers to use techniques from survival analysis. For example, the Cox proportional hazards models has been used to study factors associated with mortality using Demographic and Health Survey (DHS) or other full birth history (FBH) data \citep{pal_gender_2020,worku_survival_2011,nasejje_understanding_2015,kunnuji_background_2022,ayele_survival_2017}. Finally, \citet{okonek_pseudo-likelihood_2024} presented a pseudo-likelihood approach for estimating survival functions up to age five years using FBH data, and included a comparison of alternative parametric survival functions. However, the pseudo-likelihood approach is designed for making estimates based on one survey at a time. This paper extends \cite{okonek_pseudo-likelihood_2024} by wrapping pseudo-likelihood estimates in a Bayesian model with temporal smoothing with the ability to incorporate additional data types so that all data included by UN IGME can be utilized to produce estimates.

The outline of the paper is as follows. We start in Section \ref{sec:data} by describing the data utilized. In Section \ref{sec:models} we define the parametric survival models we consider for child mortality and then delineate the likelihood contributions of each data type, and describe the priors we select. In Section \ref{sec:results} we present two sets of results. First, we summarize pseudo-likelihood estimates using methods described by \citet{okonek_pseudo-likelihood_2024}, for available DHS and Multiple Indicator Cluster Survey (MICS) data, as well as independent yearly maximum likelihood estimates for countries with high-quality VR. We consider two parametric survival forms:~the log-logistic and piecewise-exponential models. These results help us assess the broad applicability of those two survival models across all countries. Second, we present results from our full model for four example countries with diverse data profiles (Kenya, Brazil, Estonia, and Syrian Arab Republic). These results include the application of our framework to estimate annual survival functions for the country-specific estimation years used by UN IGME, up to 2025. From the estimated survival functions, we also extracted summaries including neonatal, infant, and under-five mortality rates. Results from additional countries can be found online at \url{https://rsc.stat.washington.edu/child-survival/} and \texttt{R} code can be found at \url{https://github.com/krpaulson/childSurv}. Finally, in Section 4, we discuss extensions of this work.

\section{Data Types}\label{sec:data}

Data used for this study include all available FBH microdata from the DHS series (\url{http://www.dhsprogram.com}) and the MICS series \citep{ipums_mics_stata}; death, birth, and population counts from civil registration as prepared by UN IGME; and pre-processed estimates of neonatal, infant, or under-five mortality rate also prepared by UN IGME (available at: \url{https://childmortality.org/all-cause-mortality/data/download}). All data and estimates in our study are at the national level. We use inclusions and exclusions for each data source as determined by UN IGME to facilitate comparison of methods. The inclusion status of each observation is also available in the database at \url{https://childmortality.org/all-cause-mortality/data/download}. For each FBH survey, we choose to use a retrospective period of 20 years because estimates farther from the survey year come from the youngest mothers and may therefore be subject to bias. However, our design could be easily generalized to any other choice for the size of the retrospective period. 

\begin{figure}
    \centering
    \includegraphics[width=1\linewidth]{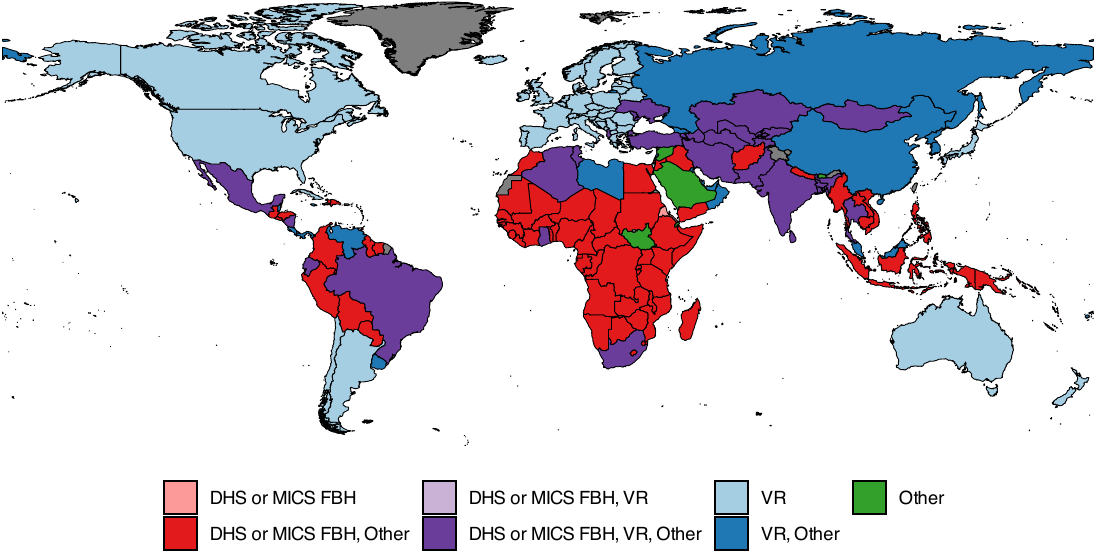}
    \caption{Countries by data type included by UN IGME. VR  = Vital registration; DHS = Demographic and Health Survey; MICS = Multiple Indicator Cluster Survey; FBH = Full birth history.}
    \label{fig:data-map}
\end{figure}

There are four types of countries UN IGME makes estimates for, and our model can accommodate any type:~$76$ countries with FBHs from DHS and/or MICS; $32$ countries with some VR in addition to DHS or MICS FBH data; $86$ countries with high-quality VR only or VR plus some non-DHS non-MICS data; and $6$ countries with no FBHs from DHS or MICS and no VR. Figure \ref{fig:data-map} shows a categorization of countries by data type included in UN IGME estimates. In general, high-income countries are most likely to have high-quality VR, while low- and middle-income countries typically rely on household survey data. There are a collection of countries -- such as Brazil, Mexico, Algeria, and many central-Asian countries -- where the UN utilizes both VR and survey data, with the VR mostly coming from more recent years. Additionally, five countries have sample vital registration (SVR) included by UN IGME: Bangladesh, China, India, Pakistan, and South Africa. SVR is classified as VR for Figure \ref{fig:data-map}. Special treatment of SVR is described in Section \ref{sec:likelihood-vr}. The six countries that do not have MICS, DHS, or VR are Bhutan, Djibouti, Micronesia (Federated States of), Saudi Arabia, South Sudan, and Syrian Arab Republic. These countries have a combination of FBHs from surveys other than DHS and MICS, summary birth histories (SBHs) from surveys or censuses, and household deaths. Countries with DHS, MICS, and/or VR also make use of additional FBH, SBH, or household deaths when available.

\section{Parametric Temporal Survival Models with Disparate Data}\label{sec:models}

\subsection{Child mortality survival model}

Let $S(a\mid \bm{\theta}_t)$ be the function identifying the probability of survival to age $a$ in months for a hypothetical child living their whole life under the mortality regime in year $t$, given parameters $\bm{\theta}_t$. In other words, $S(a\mid \bm{\theta}_t) = P(A\geq a \mid  \bm{\theta}_t)$ where $A$ is the random variable representing the age of death (in months). The survival function has a direct relationship to the mortality rates estimated by UN IGME: in units of deaths per 1000 live births, the U5MR is $1000 \times [1-S(60\mid \bm{\theta}_t)]$; the IMR is $1000 \times [1-S(12\mid \bm{\theta}_t)]$; and the NMR is $1000 \times [1-S(1\mid \bm{\theta}_t)]$.

The framework we present in this paper works for any parametric survival distribution. We discuss results for the log-logistic survival distribution and a piecewise-exponential survival distribution with breakpoints at $1$ and $12$ months. The log-logistic survival function is
$$
S(a \mid \bm{\theta}_t) =
\left[1+\left(\frac{a}{\mu_t} \right)^{1/\sigma_t} \right]^{-1}
$$
where $\mu_t>0$ and $\sigma_t>0$. We work with the transformed versions of these parameters defined as 
$$
\bm{\theta}_t = 
\begin{bmatrix}
    \theta_{t1} \\ \theta_{t2}
\end{bmatrix} =
\begin{bmatrix}
    \log(\mu_t) \\ \text{logit}(1/\sigma_t)
\end{bmatrix}.
$$
The purpose of these transformations is to impose constraints on the values of $\mu$ and $\sigma$ so that $\mu > 0$ and $\sigma > 1$. We enforce $\sigma > 1$ since this gives us a monotonically non-increasing hazard function (Appendix A). This survival model was selected based on its similarity to the log-normal model which was found by \citet{okonek_pseudo-likelihood_2024} to perform well for estimates of child mortality based on DHS data. Unlike the log-normal survival model, the log-logistic model has a tidy, closed-form solution for the logit-transformed probability of death by a given month, which facilitates the simple implementation of a normal likelihood on the logit-scale for pre-processed estimates of mortality rates, as described later in this section. Specifically, letting ${_aq_{0,t}} = 1-S(a\mid \bm{\theta}_t)$ be the probability of death by age  $a$ months in year $t$, we have,
\begin{equation}
    \text{logit}({_aq_{0,t}}) = \log \left[\left(\frac{a}{\mu_t}\right)^{1/\sigma_t} \right]
    = \text{expit}(\theta_{t2}) \left[\log a - \theta_{t1}\right].
\end{equation}
The other survival model we consider is the piecewise-exponential model with breakpoints at $1$ and $12$ months given by
$$
h(a\mid \bm{\theta}_t) = \begin{cases}
\alpha_{t1} + \alpha_{t2} + \alpha_{t3} & \text{if } a \leq 1 \\
\alpha_{t1} + \alpha_{t2} & \text{if } 1<a\leq 12 \\
\alpha_{t1} & \text{if } 12<a\leq 60
\end{cases}
$$
where $h(a\mid \bm{\theta}_t)$ is the hazard function and
$$
\bm{\theta}_t = 
\begin{bmatrix}
    \theta_{t1} \\ \theta_{t2}\\ \theta_{t3}
\end{bmatrix} =
\begin{bmatrix}
    \log(\alpha_{t1}) \\   \log(\alpha_{t2}) \\   \log(\alpha_{t3}) 
\end{bmatrix}.
$$
which ensures
$\alpha_{t1}>0$, $\alpha_{t2}>0$, $\alpha_{t3}>0$. This model assumes mortality risk is constant in the first month of life, constant at a new (lower) level for the remainder of the first year, and then constant at a third (lower) level until the fifth birthday. By using the additive formulation given here, we constrain the hazard function to be positive and non-increasing with age. Based on the piecewise-exponential model, we find the following relationship between the survival functions:
\begin{align}
    S(1\mid \bm{\theta}) &= \exp(-[\alpha_1 + \alpha_2 + \alpha_3]) \\
    S(12\mid \bm{\theta}) &= \exp(-[12 \alpha_1 + 12 \alpha_2 + \alpha_3]) \\
    S(60\mid \bm{\theta}) &= \exp(-[60 \alpha_1 + 12 \alpha_2 + \alpha_3]).
\end{align}
While the 2-parameter log-logistic model is more parsimonious than the 3-parameter piecewise-exponential model, the piecewise-exponential model is more flexible, specifically with respect to values of the survival function at the breakpoints. Apart from the non-decreasing relationship enforced by the definition of a survival function, the piecewise-exponential model imposes no structure on the relationship between NMR, IMR and U5MR.

We also compare our findings based on FBH data to the commonly-used discrete hazards model as described by \cite{mercer2015space}. The discrete hazards model is a survival model, but not a continuous one like the log-logistic and piecewise-exponential described above. Rather, survival probabilities are defined for each discrete month of age. The discrete hazards model is used by UN IGME for small-area estimates \citep{wu2021dhs} and is also the model used by the DHS for generating U5MR estimates in their official reports \citep{dhs-guide}. For VR count data, we include comparisons to the life table method described by \cite{explanatory-igme}.

\subsection{Bayesian hierarchical model for child survival}

We model each country separately rather than in a global model. The following holds for a generic country and the notation omits a country index for simplicity. We aim to make estimates for each year in a country-specific set, starting with the first year estimated by UN IGME for the particular country and ending with 2025. We index these years by $t \in \{1,...,T\}$. Let $\bm{y} := (\bm{y}_1, \ldots, \bm{y}_T)$ be the totality of the data for a country and $\bm{\lambda}$ be a set of hyperparameters. Take $\bm{\theta}$ to be the $T \times 2$ or $T \times 3$ matrix of survival parameters where row corresponds to year and column corresponds to the distinct survival parameters (i.e. the two parameters for the log-logistic or the three for the piecewise-exponential). The Bayesian model is
\begin{align}
\bm{y} \mid  \bm{\theta} &\sim p_1(\bm{y} \mid \bm{\theta}) \\
\bm{\theta} \mid  \bm{\lambda} &\sim p_2(\bm{\theta}\mid \bm{\lambda}) \\
\bm{\lambda} &\sim p_3(\bm{\lambda})
\end{align}
where $p_1$ is the likelihood of the data given the survival parameters; $p_2$ is the prior for $\bm{\theta}$ given hyperparameters $\bm{\lambda}$, which utilizes a multivariate random walk temporal structure; and $p_3$ is the hyperprior for $\bm{\lambda}$. We use this model to make inference on $\bm{\theta}$, and we use the posterior samples for $\bm{\theta}$ to construct estimates of child mortality rates for the age groups of interest. In the remainder of this section we describe each component of this model in detail.

\subsubsection{Likelihood for FBH microdata}\label{sec:likelihood-fbh}

As previously noted, there are multiple distinct types of data available for mortality estimation and each contributes to the joint likelihood in its own way. When microdata from FBHs are available, they include observations of age at death or age at interview for children of surveyed mothers. From a single survey $s$ conducted in year $t$, we have information on childhood survival for the previous 20 years data, due to the retrospective nature of the data -- mothers between the ages of 15 and 49 are asked for their complete birth histories, including any deaths to children. Hence, from survey $s$ we obtain estimates of survival parameters for 20 years, denoted $\begin{bmatrix}\hat{\bm{\theta}}_{t-19,s} ^\top \cdots \hat{\bm{\theta}}_{t,s}\end{bmatrix}^\top$. We use pseudo-likelihood estimation which correctly accounts for the survey design \citep{binder1983variances, okonek_pseudo-likelihood_2024}. As described by \citet{okonek_pseudo-likelihood_2024}, we use interval censoring to address potential age-heaping in survey data at 12 months of age; children recorded to have died at 12 months are interval censored to the period between 6 and 18 months of age. Pseudo-likelihood estimation is completed using the \texttt{pssst} package in \texttt{R} \citep{pssst}. 

The likelihood contribution of survey $s$ to our full model is
\begin{align}
\begin{bmatrix} \widehat{\bm{\theta}}_{t-19,s} \\ \vdots \\ \widehat{\bm{\theta}}_{t,s} \end{bmatrix} \Bigg|
\begin{bmatrix} \bm{\theta}_{t-19} \\ \vdots \\ \bm{\theta}_{t} \end{bmatrix}, \bm{\lambda}
    \sim \text{N}\left(\begin{bmatrix} \bm{\theta}_{t-19} \\ \vdots \\ \bm{\theta}_{t} \end{bmatrix}, \widehat{\bm{V}}_s\right)
\end{align}
where $\widehat{\bm{V}}_s$ is the variance-covariance matrix for the pseudo-likelihood estimate. Note that multiple surveys can contribute to the estimates of $\bm{\theta}_t$ in any year $t$. For this paper, we use microdata from DHS and MICS FBHs only. However, any FBH with microdata could be processed in this way.

One limitation of FBH data is that mothers who died of HIV/AIDS are unavailable to be surveyed, and their children are also more likely to have died, potentially inducing downward bias in derived child mortality rates. For countries with high HIV/AIDS burden, UN IGME uses an adjustment to U5MR to address this bias in FBH data \citep{explanatory-igme,walker_child_2012}. UN IGME applies no adjustment to NMR or to IMR. We use the same U5MR adjustment factors as UN IGME, and translate the adjustment to the survival model using the methods described in Appendix B.

\subsubsection{Likelihood for death counts from complete VR}\label{sec:likelihood-vr}

For death counts from complete VR, the first step is to reformat available data for each country-year to generate non-overlapping age groups, so that we can create independent likelihood contributions. Use ${_nD_{a,t}}$ to denote the observed number of deaths between ages $a$ and $a+n$ in months, in year $t$. For neonatal deaths in year $t$, the likelihood contribution is Poisson-lognormal according to
\begin{align}
{_1D_{0,t}}\mid \bm{\theta}_t, \bm{\lambda} &\sim \text{Poisson}(B_t \times [1-S(1\mid \bm{\theta}_t)] \times \exp(\kappa_{1,t})) \\
\kappa_{1,t} \mid \phi &\sim \text{N}(0, \phi^{-1})
\end{align}
where $B_t$ is the number of births in year $t$ and $\phi$ is a parameter that captures overdispersion, since it is the precision of the random effects $\kappa_{1,t}$. For death counts from ages 1, 2, 3, or 4 years; 1-4 years; or total under-5 years, the likelihood contribution is
\begin{align}
{_nD_{a,t}}\mid \bm{\theta}_t, \bm{\lambda} &\sim \text{Poisson}({_nm_{a,t}} \times {_nP_{a,t}} \times \exp(\kappa_{a,t})) \\
\kappa_{a,t} \mid \phi &\sim \text{N}(0, \phi^{-1})
\end{align}
where
\begin{align}
    {_nm_{a,t}} = \frac{12 \times [S(a\mid \bm{\theta}_t) - S(a+n \mid \bm{\theta}_t)]}{\int_{a}^{a+n} S(x\mid \bm{\theta}_t)
    ~dx}
\end{align}
is the mortality rate in units of deaths per person-year in year $t$, ${_nP_{a,t}}$ is the mid-year population between ages $a$ and $a+n$, in year $t$, and $\kappa_{a,t}$ is the log-normal random effect for age group $a$ and year $t$. The constant 12 in the numerator is used to convert deaths per person-months into deaths per person-year---a necessary conversion because the mid-year population is an approximation for person-years of exposure time.

For post-neonatal death counts (1 to 11 months of age), we do not have access to the mid-year post-neonatal population, but rather the number of births and mid-year infant population. Since ${_{11}D_{1,t}} = {_{12}D_{0,t}} - {_1D_{0,t}}$ (post-neonatal deaths equals infant deaths minus neonatal deaths) we can use the form
\begin{align}
{_{11}D_{1,t}}\mid \bm{\theta}_t, \bm{\lambda} &\sim \text{Poisson}\left( \left\{ {_{12}m_{0,t}} \times {_{12}P_{0,t}} -B_t \times [1-S(1\mid \bm{\theta}_t)] \right\} \exp(\kappa_{a,t}) 
\right)
\\
\kappa_{a,t} \mid \phi &\sim \text{N}(0, \phi^{-1}).
\end{align}
Since we have chosen the age groups to be non-overlapping so that the corresponding death counts are independent conditional on $\bm{\theta}_t$, the joint likelihood for all death counts from VR is the product of these age-specific overdispersed-Poisson likelihoods. With full VR data (i.e.,~all deaths measured), we might ask where the stochastic variation is arising from. Underlying the empirical death counts is a generative model and since we are interested in temporal trends, it is this that we are attempting to recover with our model. In any given year, there is sampling variation around the generative mean, and it is this variation that we are smoothing over.

For sample vital registration, if counts are available, the above likelihoods can be adapted by applying the fraction of the total population sampled as a scalar to $B_t$ and ${_nP_{a,t}}$. There are also some cases where VR or SVR data are available but only in a report listing one or more mortality rates and not in count space. For these cases, we use a Poisson approximation to estimate a standard error for the pre-processed mortality rate and use the likelihood as described in Section \ref{sec:likelihood-pp}.

\subsubsection{Likelihood for pre-processed estimates of mortality rates}\label{sec:likelihood-pp}

In some cases, reports are available with pre-processed mortality rates with a standard error, but FBH microdata or VR death counts are not available. We also include FBHs which are not DHS or MICS in this category, but future work could be done to process all FBH with microdata according to the methods described in Section \ref{sec:likelihood-fbh}. Let the $i$-th observation of this kind corresponding to probability of death before the $a$-th month and to children in year $t$, be denoted ${_aq_{0,i,t}}$. For these observations, the likelihood contribution is
\begin{align}
\text{logit}({_aq_{0,i,t}}) \mid \bm{\theta}_t, \bm{\lambda} \sim \text{N}(\text{logit}[1-S(a\mid \bm{\theta}_t)], \nu_{i,a,t}),
\end{align}
where $\nu_{i,a,t}$ be the corresponding reported sample variance (obtained via the delta method if the reported variance is for ${_aq_{0,i,t}}$).
Summary birth histories (SBHs) are also incorporated in this way. The IMR and the U5MR are estimated from each available SBH and these estimates contribute logit-normal likelihoods. The sampling variance we use for SBH is described in Appendix C.

\subsubsection{Summary of joint likelihood}

In summary, let the observations for a given country be the data vector $\bm{y} =(\widehat{\bm{\theta}}, \bm{D}, \text{logit}(\bm{q}))$, where $\bm{\widehat{\theta}}$ is the vector of pseudo-likelihood estimates for $\bm{\theta}$ based on FBH microdata, $\bm{D}$ is the vector of observed death counts, and $\bm{q}$ is the vector of pre-processed estimates of child mortality rates. The joint-likelihood is,
\begin{align}
p_1(\bm{y} \mid  \bm{\theta}, \bm{\lambda}) = 
p(\widehat{\bm{\theta}} \mid \bm{\theta}, \bm{\lambda})
p(\bm{D} \mid  \bm{\theta}, \bm{\lambda})
p(\text{logit}(\bm{q}) \mid  \bm{\theta}, \bm{\lambda}).
\end{align}
We would also have associated variance terms, and these would be conditioned upon in $p_1$, but for simplicity, we have left such terms out. This joint likelihood assumes the observations from different sources are independent conditional on $\bm{\theta}$. This is a reasonable assumption since distinct surveys will come from independent samples and survey data is not usually used for years with available VR data.

\subsubsection{Priors, hyperpriors, and computation}

We now consider the prior for the parameter matrix $\bm{\theta}$. We would like to encourage temporal smoothness which allows us to leverage information across years. We choose separate second-order random walk priors for each of the two or three survival parameters. Without loss of generality, let $\theta_{t}$ for $t\in \{1,\ldots,T\}$ be the values of an arbitrary parameter (i.e. one of the two or three depending on the survival model). We choose the model
$$
\theta_t = \beta + \delta_t + \varepsilon_t,
$$
where $\beta$ is a parameter-specific intercept, $\delta_t$ is the term for year $t$ that comes from a random walk, and $\varepsilon_t$ represents unstructured temporal variation in the observations not captured by the random walk. The intercept is given a normal prior based on the global range of parameter values coming from either pseudo-likelihood estimates or VR maximum likelihood estimates. The form of the random walk is
$$
(\delta_t - \delta_{t-1}) - (\delta_{t-1} - \delta_{t-2}) \mid \tau_\delta \sim
\text{N}(0, \tau_\delta^{-1})
$$
for $t=3, \dots, T$. The unstructured terms are independent and identically distributed (i.i.d.) with respect to time with $\varepsilon_1, \ldots, \varepsilon_T \mid \tau_\varepsilon \overset{\text{i.i.d.}}{\sim} \text{N}(0, \tau_\varepsilon^{-1})$.

We use penalized complexity (PC) priors for the precision in both the random walk and the unstructured temporal term \citep{simpson_penalising_2017}. Finally, we use Template Model Builder (TMB) for computation, as its use of Laplace approximations makes it computationally efficient, and unlike the integrated nested Laplace approximation (INLA) approach \citep{rue_bayesian_2017} it is flexible enough to accommodate the non-linear form of this model \citep{tmb}. Code for the TMB model is available at \url{https://github.com/krpaulson/childSurv}. TMB has been shown to be accurate in a far-reaching simulation exercise and comparison with INLA \citep{osgood2023statistical}. Computing annual pseudo-likelihood estimates using the \texttt{pssst} \texttt{R} package \citep{pssst} can take several hours, and fitting our full model in TMB has a typical computation time on the order of minutes.

\section{Results}\label{sec:results}

\begin{figure}[ht]
    \centering
    \begin{subfigure}{0.8\textwidth}
        \centering
        \includegraphics[width=\linewidth]{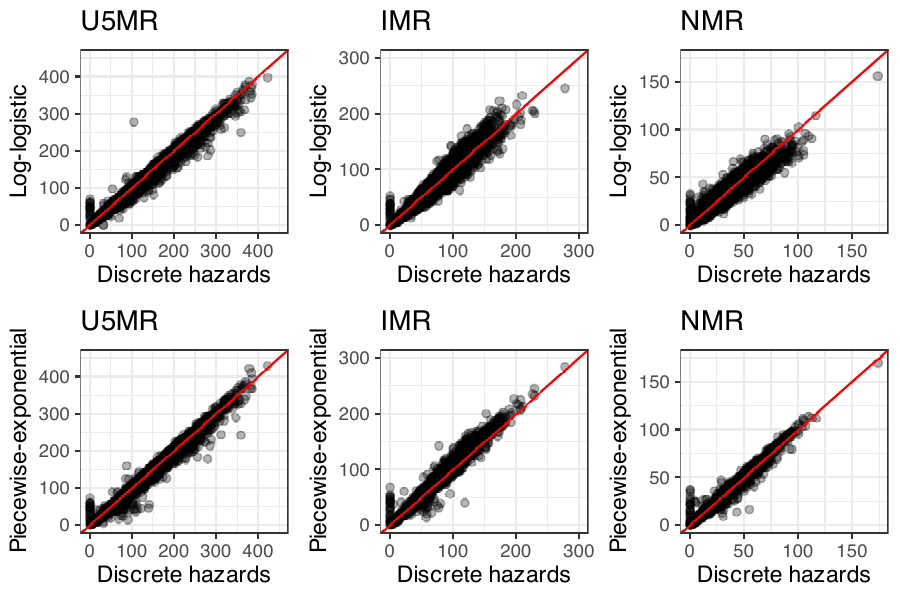}
        \caption{Survey}
    \vspace{1em}
    \end{subfigure}
    
    \begin{subfigure}{0.8\textwidth}
        \centering
        \includegraphics[width=\linewidth]{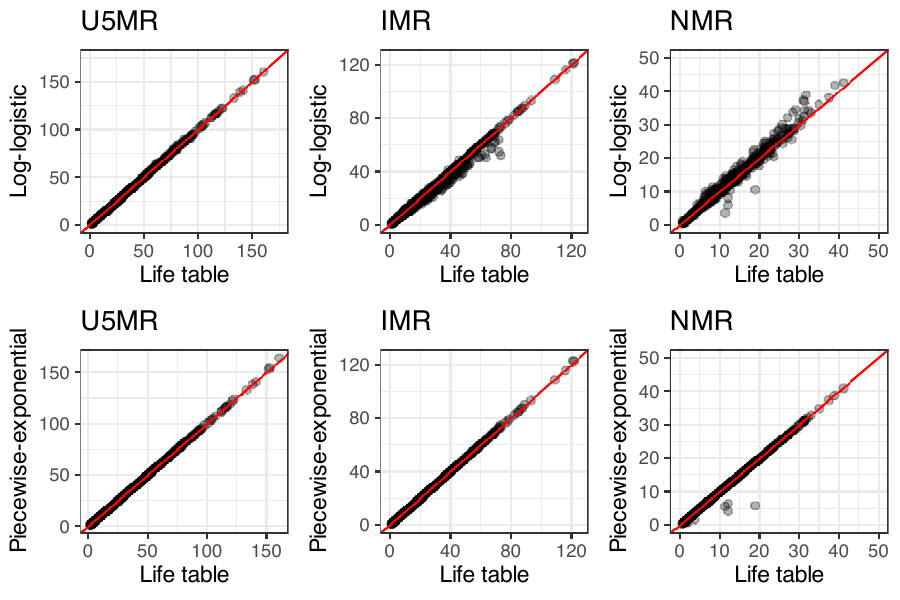}
        \caption{Vital Registration}
    \end{subfigure}
    \caption{Under-five mortality rate (U5MR), infant mortality rate (IMR), and neonatal mortality rate (NMR) from (a) survey data using pseudo-likelihood estimation and from (b) vital registration data using maximum likelihood estimation, according to the  log-logistic survival  and piecewise-exponential models. Comparison for survey data is against the discrete hazards model and comparison for vital registration is against the life table method used by UN IGME to convert counts to probabilities of death. The red diagonal line is a line of equivalence.}
    \label{fig:scatters}
\end{figure}

Results for annual estimates of NMR, IMR, and U5MR -- arising from pseudo-likelihood estimation for available DHS and MICS and maximum-likelihood estimation for VR counts -- are presented in Figure \ref{fig:scatters}. We compare our estimates based on the log-logistic and piecewise-exponential models to the current best practice methods -- the discrete hazards method for survey data \citep{mercer2015space} and a life table based approach for VR data \citep{explanatory-igme}. Country-specific versions of these plots for our example countries can be found in Appendix D. Our results suggest that both the log-logistic and the piecewise-exponential models have reasonable agreement with current best-practice on average. However, there are some cases where the log-logistic model has systematic difference relative to the discrete hazards or life table approach. For example, IMR for Brazil appears to be under-estimated by the log-logistic model but not the piecewise-exponential model (Figure \ref{fig:brazil-comp-to-dh}). Among VR countries, we see that the piecewise-exponential model has very strong alignment with the life table method, whereas the log-logistic model performs very well for many countries but does have more deviations from the life table method than the piecewise-exponential does for IMR and NMR in particular, see Figure \ref{fig:scatters} (b). Generally speaking, we observe more variability on the basis of model choice from survey data than from VR data. This is probably due to the extra complexity of survey data and additional modeling choices we have made such as our approach to interval censoring for age heaping (described in Section \ref{sec:likelihood-fbh}). A summary of the estimated survival parameter values is included in Appendix D (Figure \ref{fig:par-ests-all}).

\begin{figure}
    \centering
    \includegraphics[width=1\linewidth]{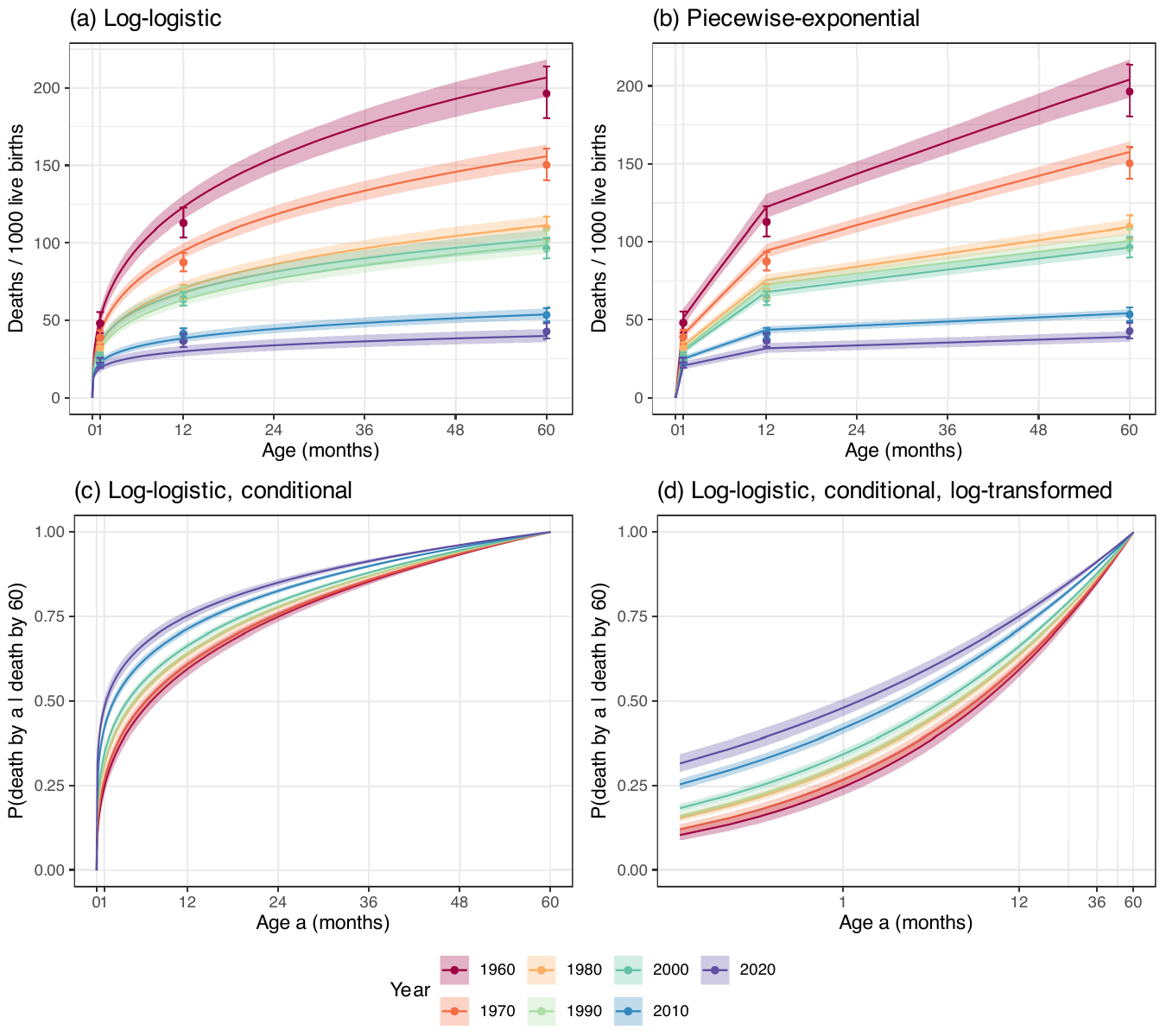}
    \caption{Posterior median and 90\% intervals for deaths per 1000 live births before each month of age, up to the fifth birthday, for Kenya, according to the (a) log-logistic and (b) piecewise-exponential survival models. Deaths per 1000 live births is equal to $1000 \times \Pr ( A \leq a )=1000\times [1-S( a | \bm{\theta})]$ where $A$ is the random variable for time at death in months. Estimates of the NMR, IMR, and U5MR from UN IGME are indicated with points and error bars at 1, 12, and 60 months, respectively, representing 90\% intervals. The figure also includes (c) the probability of death by $a$ months of age conditional on death by 60 months of age, and (d) the same thing but with a log base 2 transformation applied to age, again with posterior median and 90\% intervals. Results for all panels are presented for the years 1960, 1970, 1980, 1990, 2000, 2010, and 2020.}
    \label{fig:curves}
\end{figure}

Based on the results from our model, we can extract estimates of the probability of death in any subset of the under-five age group, for any country and year. Mortality rates such as NMR, IMR, and U5MR are probabilities of death between birth and age 1, 12, or 60 months, respectively. Figures \ref{fig:curves}a and \ref{fig:curves}b show the posterior median and 90\% credible interval for the probability of death as a function of age, in units of deaths per 1000 live births, for Kenya, according to the log-logistic and piecewise-exponential models. Our results show reasonable alignment with the UN IGME 2024 estimates for NMR, IMR, and U5MR, but also illustrate the extra information produced by our model, since we have obtained a full curve and not just three points for each country-year. While probability of death between birth and a given month of age is displayed in Figure \ref{fig:curves}a and Figure \ref{fig:curves}b, we emphasize that we can use the results of our model to derive the probability of death between age $a$ and $a+n$ for any $a$ and $a+n$ contained within the under-five period. Figures \ref{fig:curves}a and \ref{fig:curves}b also illustrate that while the piecewise-exponential model may be more flexible to align with data for NMR, IMR, and U5MR specifically, the shape of the curve is perhaps less realistic than that produced by the log-logistic model because of the constant hazard assumption.

Our model also allows for valid inference on any derivative of the estimated survival functions, including the probability of death conditional on death by the fifth birthday (60 months of age). This particular probability is important because it can tell us about the relative change over time when comparing different ages in the under-five age group. For instance, we find that in Kenya, the probability of an under-five death occurring in the neonatal period was $0.25$ (90\% CI $[0.22, 0.27]$) in 1960 and $0.48$ (90\% CI $[0.46, 0.51]$) by 2020 (Figures \ref{fig:curves}c and \ref{fig:curves}d). Hence, the shape of the survival curve has changed over time; while mortality rates have declined in every age group between 1960 and 2020, there has been much greater progress in the older age groups than in neonatal mortality.

\begin{figure}
    \centering
    \includegraphics[width=0.9\linewidth]{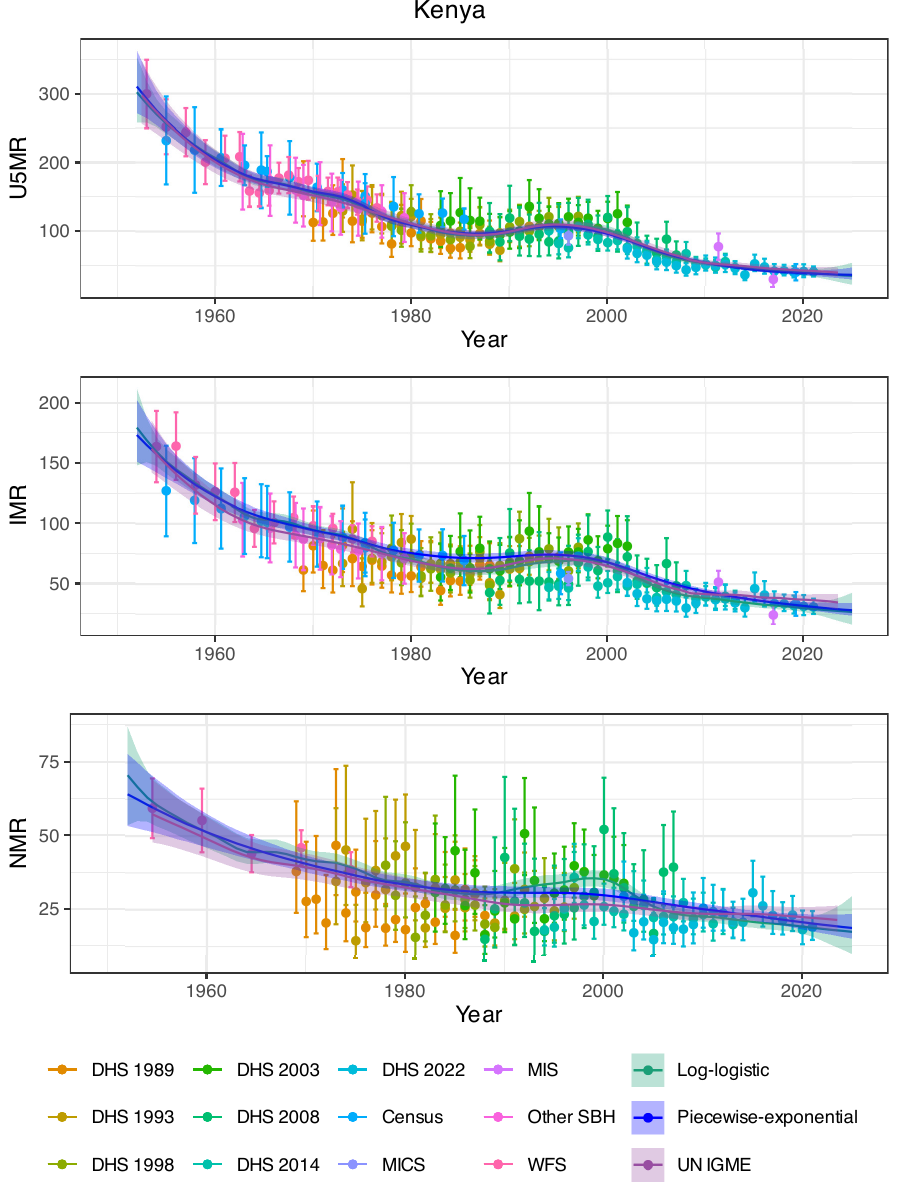}
    \caption{Posterior median and 90\% credible intervals for estimates of the U5MR, IMR, and NMR in Kenya according to the log-logistic and piecewise-exponential survival models. The UN IGME 2024 estimates and 90\% intervals are also included for reference. Input data are plotted with point estimates and 95\% error bars. For DHS full birth histories, the input data presented come from the discrete hazards model. DHS = Demographic and Health Survey; MICS = Multiple Indicator Cluster Survey; SBH = summary birth history; WFS = World Fertility Survey; MIS = Malaria Indicator Survey; U5MR = under-five mortality rate; IMR = infant mortality rate; NMR = neonatal mortality rate.}
    \label{fig:countries-kenya}
\end{figure}

\begin{figure}
    \centering
    \includegraphics[width=0.9\linewidth]{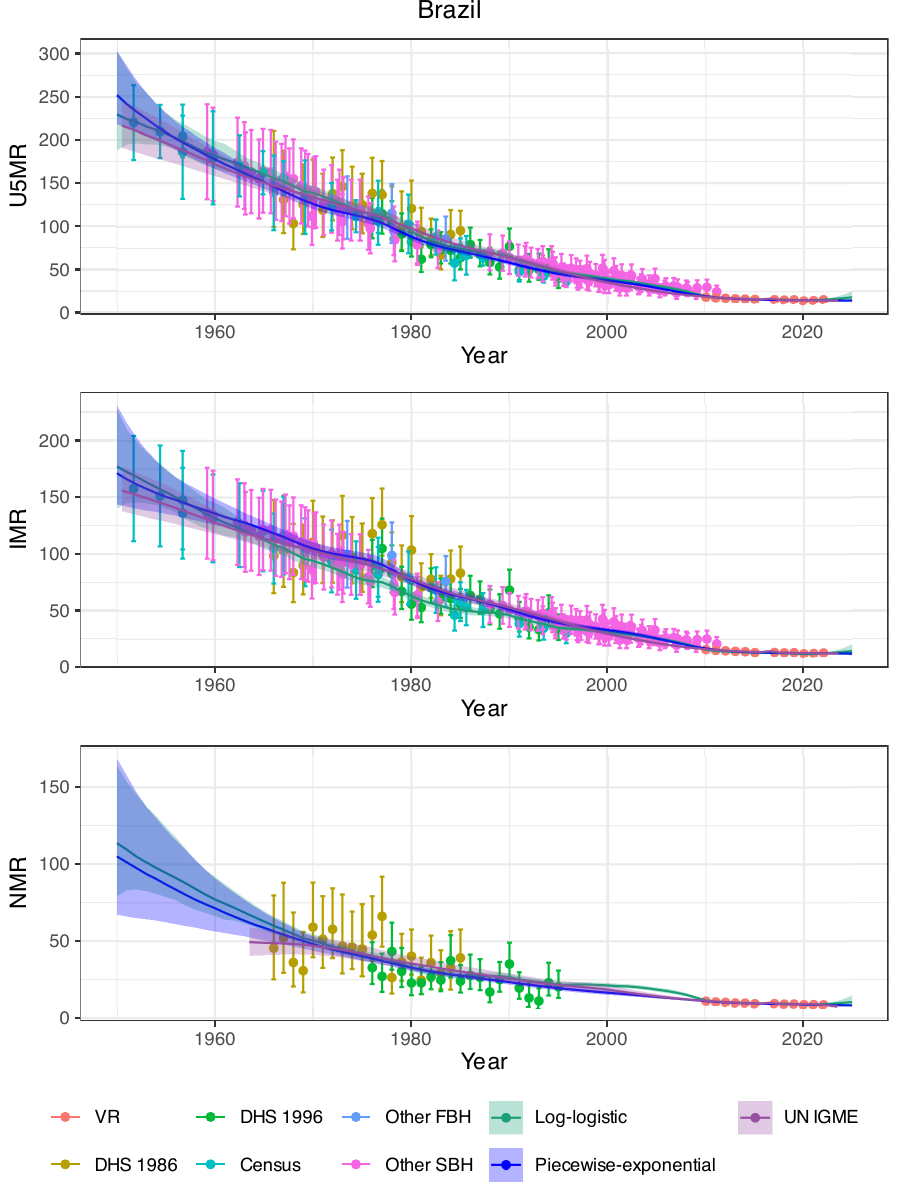}
    \caption{Posterior median and 90\% credible intervals for estimates of the U5MR, IMR, and NMR in Brazil according to the log-logistic and piecewise-exponential survival models. The UN IGME 2024 estimates and 90\% intervals are also included for reference. Input data are plotted with point estimates and 95\% error bars. For DHS full birth histories, the input data presented come from the discrete hazards model. DHS = Demographic and Health Survey; SBH = summary birth history; VR = Vital Registration; FBH = full birth history; U5MR = under-five mortality rate; IMR = infant mortality rate; NMR = neonatal mortality rate.}
    \label{fig:countries-brazil}
\end{figure}

\begin{figure}
    \centering
    \includegraphics[width=0.9\linewidth]{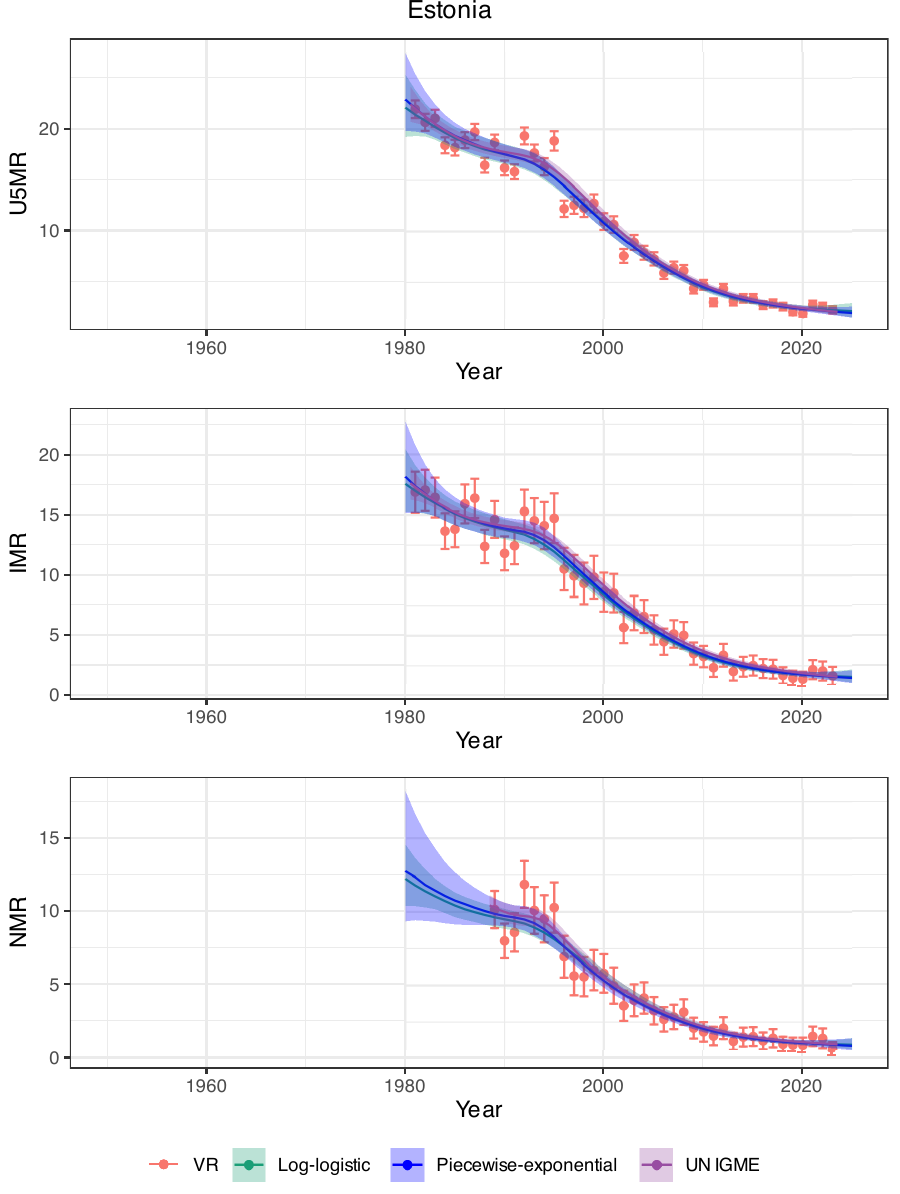}
    \caption{Posterior median and 90\% credible intervals for estimates of the U5MR, IMR, and NMR in Estonia according to the log-logistic and piecewise-exponential survival models. The UN IGME 2024 estimates and 90\% intervals are also included for reference. Input data are plotted with point estimates and 95\% error bars. For DHS full birth histories, the input data presented come from the discrete hazards model. VR = Vital Registration; U5MR = under-five mortality rate; IMR = infant mortality rate; NMR = neonatal mortality rate.}
    \label{fig:countries-estonia}
\end{figure}

To illustrate the flexibility of our model to accommodate countries with different data profiles, we present findings for Kenya, Brazil, Estonia, and Syrian Arab Republic. Estimates of the U5MR in these countries over time, according to both the log-logistic and the piecewise-exponential survival models, are presented in Figures \ref{fig:countries-kenya}, \ref{fig:countries-brazil}, \ref{fig:countries-estonia}, and \ref{fig:countries-syria}. Plots of the estimated survival parameters over time, are included in Appendix D.

Kenya is an example of a country with strong coverage of DHS data, in addition to some SBH data. It is also one of the 17 countries where UN IGME applies the missing mothers adjustment for HIV. Our results for the U5MR in Kenya for both survival models agree strongly with the UN IGME estimates (Figure \ref{fig:countries-kenya}). The IMR is also largely in agreement across our two models and between our models and the UN IGME estimates. However, the IMR estimated by the piecewise-exponential model is somewhat higher than the IMR from the log-logistic model after 1975. In the case of the NMR, we observe some separation between the models in the period most impacted by HIV. Part of this is the impact of our missing mothers adjustment, which applies the adjustment evenly across age, as compared to the UN IGME missing mothers adjustment, which is not applied at all to the neonatal period. We also estimate a stronger pattern of increasing followed by decreasing NMR by the log-logistic model than the piecewise-exponential model. In this case, the log-logistic model propagates patterns observed in older children to the NMR whereas the piecewise-exponential model has sufficient flexibility to allow for very different temporal trends across the three indicators.

Input data included for Brazil are two DHS surveys from 1986 and 1996, census-based SBH, SBH from Pesquisa Nacional por Amostra de Domicilios, SBH from the Brazil National Health Survey, FBH from Pesquisa Nacional por Amostra de Domicilios 1986, and VR between 2010 and 2022 (Figure \ref{fig:countries-brazil}). As opposed to death counts, the VR we use for Brazil (in alignment with UN IGME) are the completeness-adjusted mortality rates which are the official rates published by the Brazil Ministry of Health. The results for Brazil are one example of successful integration of survey and vital registration data by our model to produce one cohesive set of estimates over time. As with Kenya, our models produce estimates for Brazil in broad agreement with the UN IGME estimates, especially for the U5MR. The IMR is somewhat under-estimated by the log-logistic model, relative to the discrete hazards and piecewise-exponential models (Figures \ref{fig:countries-brazil} and \ref{fig:brazil-comp-to-dh}).

For Estonia, we incorporate annual death counts from VR only, hence this example illustrates that our model framework works well for high-quality VR countries in addition to survey-based LMICs. There is strong agreement between our two survival models and the UN IGME 2024 estimates for all three primary indicators (Figure \ref{fig:countries-estonia}). Since Estonia is a country with a relatively small population, there is visible stochastic noise in the observations, and our results for the mean mortality risk show satisfactory smoothing over time. The UN vital registration database has counts of infant and under-five deaths back to 1981, but neonatal death counts only back to 1989. Hence, we can see in our results one of the benefits of assuming a survival model with more structure (e.g.,~the log-logistic model) as compared to a more flexible model (e.g.,~the piecewise-exponential model); both of our models are able to produce estimates of the NMR back in time, but the more structured log-logistic model generates smaller uncertainty intervals for the NMR for years with no neonatal data (Figure \ref{fig:countries-estonia}).

Finally, Syrian Arab Republic is an example of a country with no DHS or MICS full birth histories and no VR. Therefore, all data for this country as incorporated into our model are of the pre-processed estimate type. For someone interested in applying these methods to this country in particular, an extension of our work could be to get access to the microdata for the Pan-Arab Project for Family Health FBHs and process those data in the manner we have described for DHS and MICS. One difference between our estimates and the UN IGME estimates is that we have not included shocks from conflict, natural disasters, and other short-term events (Figures \ref{fig:countries-syria}). We suggest using the same increase in the U5MR as UN IGME and adopting a similar approach to our missing mothers adjustment to apply the inflation evenly over age (Appendix B). Another phenomenon we see for Syrian Arab Republic is that the log-logistic model estimates that the World Fertility Survey NMR data from before 1980 is too low. Instead, it produces a fit which better matches the universe of expected relationships between the NMR and the U5MR (Figure \ref{fig:syria-nmr-u5mr-scatter}). Conversely, the piecewise-exponential model has much more flexibility for the ratio of NMR and U5MR to be whatever is indicated by the country-specific data, and we see evidence of this in our results for Syrian Arab Republic. Lastly, the most recently included data for this country is from 2008, meaning that we are asking our model to make projections for 17 years to reach 2025. In cases like this, we caution against over-interpretation of the posterior median and instead suggest consideration of the credible interval when using or discussing these results.

\begin{figure}
    \centering
    \includegraphics[width=0.9\linewidth]{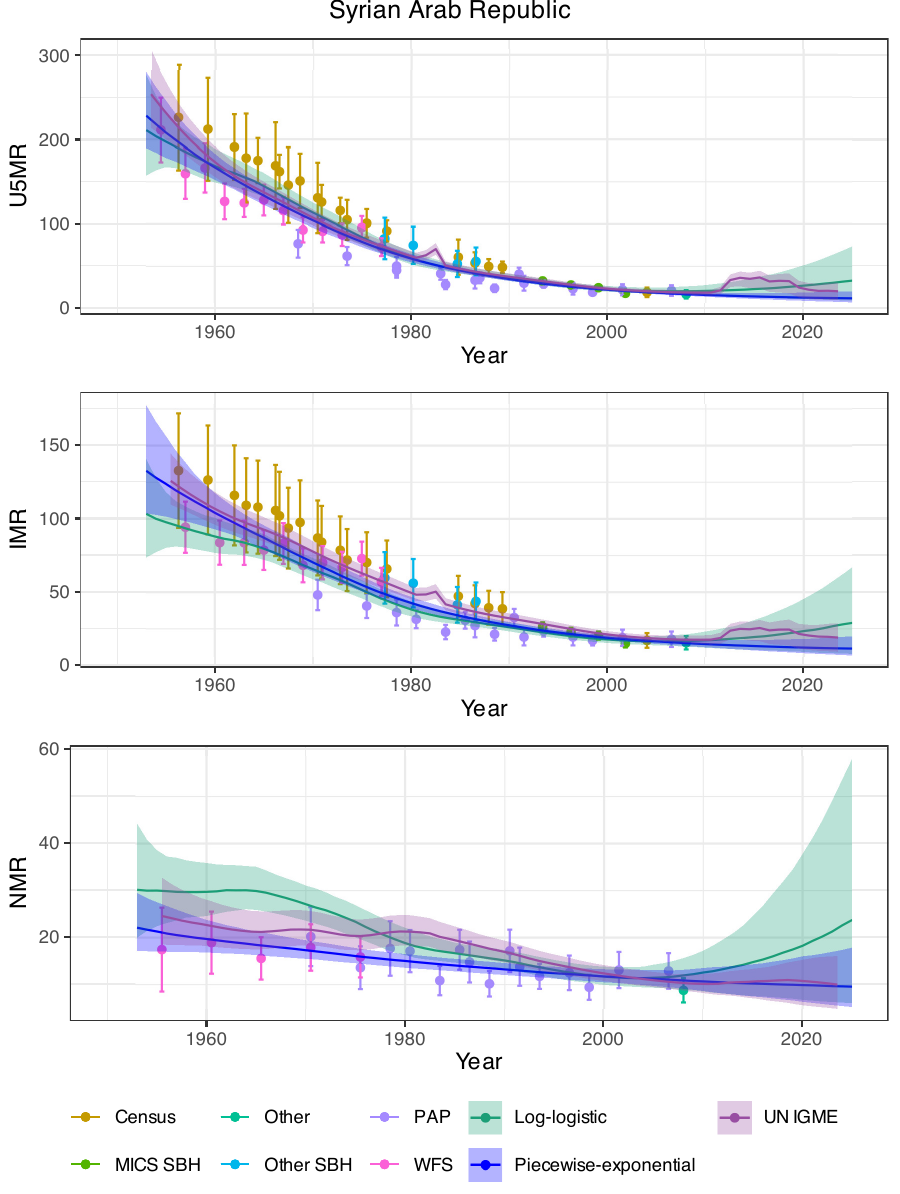}
    \caption{Posterior median and 90\% credible intervals for estimates of the U5MR, IMR, and NMR in Syrian Arab Republic according to the log-logistic and piecewise-exponential survival models. The UN IGME 2024 estimates and 90\% intervals are also included for reference. Input data are plotted with point estimates and 95\% error bars. For DHS full birth histories, the input data presented come from the discrete hazards model. MICS = Multiple Indicator Cluster Survey; SBH = summary birth history; WFS = World Fertility Survey; PAP = Pan-Arab Project for Family Health; U5MR = under-five mortality rate; IMR = infant mortality rate; NMR = neonatal mortality rate.}
    \label{fig:countries-syria}
\end{figure}

\section{Discussion}\label{sec:discussion}

In this paper we have demonstrated that a survival framework can be used to estimate child mortality rates at the national level using VR death counts, full birth history microdata, and/or reported point estimates of child mortality rates. While survival models have been used before for analyzing a single data type, e.g.,~full birth history data \citep{okonek_pseudo-likelihood_2024}, our work is novel for the integration of all data types available for the estimation of child mortality rates.

To further improve upon the work we have presented here, we suggest the following extensions. First, we find that the log-logistic survival model is useful because it has relatively few parameters, and is therefore computationally simple and can produce efficient inference when it fits the data well. The log-logistic model fits reasonably well for the majority of countries. However, there are some countries where the log-logistic model is not sufficiently flexible, and we find significant deviations from the discrete hazards or life table point estimates for U5MR, IMR, and/or NMR. The piecewise-exponential survival function with three levels -- one for the first month, one for the remainder of the first year, and one for ages 1-4 years -- is more flexible but fitting three parameters is a larger computational challenge and the model may not provide enough structure for cases such as VR country-years with infant and under-five counts but no neonatal counts.

To decide between these two survival models in practice, we recommend comparing estimates of key summaries, such as the NMR and U5MR, with estimates from discrete hazards and life table approaches. The availability of neonatal death counts for VR countries is also important since if absent there is an identifiability issue for the piecewise exponential model. One can also assess whether the extra structure provided by the log-logistic model might be beneficial, as it was for the Syrian Arab Republic, and also whether the log-logistic model is propagating temporal trends from U5MR to other age groups and if this is desirable (as for  Kenya). All else being equal, the log-logistic model should be preferred because it has fewer parameters. Ideally, an application of the general framework presented in this paper would be based on a single survival model that works well in all countries, therefore eliminating subjective model selection, and we plan on pursuing this objective. An extension of this study therefore is to further explore other existing parametric survival functions in the context of child mortality estimation, or to propose a new survival function. One alternative is a monotonic non-increasing spline on the hazard function with knots at, for example, 1 and 12 months. A second alternative is a mixture model.

Second, the temporal model produces adequate estimates for country-years with data or country-years between observations, but may not always produce useful estimates for forecasts beyond a few years. The B3 model used by UN IGME makes projections which are an ad hoc weighted average of global and national trends. This approach occasionally results in declining forecasts which do not appear to be supported by location-specific data. However, there is an opportunity to develop a new forecasting method which is better for child mortality than a simple intrinsic temporal model like the random walk used here \citep{paulson-2025}. This issue may be especially important in the future given the current uncertainty around the fate of the DHS program. However, we emphasize that no forecasting model can make up for a lack of data, particularly in a landscape where child mortality rates may very well sharply change as a result of a reduction in funding for global health programs. Extensions of this study may also seek to address correlation between the survival parameters, such as by using a multivariate random walk as described by \citet{okonek_dissertation}.

Third, our model could be extended to include corrections for bias from various sources. For example, survey-specific bias terms could be introduced to account for systematic error in particular surveys. We present one method for making an HIV adjustment to account for missing mothers in survey data, based on the U5MR adjustments made by UN IGME \citep{explanatory-igme, walker_child_2012}. The adjustment needs to be made on the observations, so we need to apply it to the pseudo-likelihood estimates of the survival parameters directly. We adapt the missing mothers adjustment to the survival model by assuming the bias is distributed approximately equally over age on a multiplicative scale to the hazard function. However, current convention is to apply the adjustment to post-neonatal age groups only, and it would be impossible to achieve this while maintaining the same functional form for survival unless a piecewise-exponential hazard is used. Adaptations of this paper could explore other approaches to this HIV correction. Next, the B3 model includes a feature that allows for bias to be different for data points corresponding to longer retrospective periods relative to the survey year. One rationale for this is that data on births that occurred further in the past are subject to the most recall bias and survivor bias, and will also correspond to children born to the youngest mothers at the time of birth. However, there is limited literature on the consistency, strength, or direction of these potential biases. The B3 approach to bias correction could be re-evaluated and either used directly or modified for use in this model. Finally, summary birth history data could be carefully studied for differential bias by age group of the mother or time since first birth \citep{verhulst_child_2016}.

Fourth, further model development to include terms for additional data that is currently excluded by UN IGME would mark a significant improvement. \cite{alkema2014global} discuss a method for including incomplete VR, but it is not currently utilized by UN IGME. One potential way to include incomplete VR would be to model completeness as linear (on some scale, for example, the logistic) with respect to time, and then use the country-time-specific completeness as a multiplier on the mean of the Poisson-lognormal likelihood for death counts. VR and non-VR sources alike are frequently excluded by UN IGME based on expert opinion and visual comparison to previously collected data. This method is ad hoc and not as evidence-based as desirable. A method to include all available data while down-weighting in some manner the observations that are further from the other observations would be desirable.

In summary, we have outlined a flexible survival framework for estimating child mortality rates at the national level, while incorporating the different data types commonly found for child mortality. We have produced code to fit these models using TMB and demonstrated that we can achieve reasonable, precise results for a large number of countries. This framework could easily be adapted to assume a different family for the underlying survival model, or, it could be modified to use a different latent temporal model. The framework could also be extended to make estimates of age-specific mortality beyond age five, although this may require a different survival model. A survival analysis approach is appropriate for this application and represents an improvement over the multi-model approach currently used by UN IGME and IHME for child mortality estimation at the national level.

\FloatBarrier

\bibliographystyle{plainnat}
\bibliography{references}

\clearpage
\newpage

\setcounter{figure}{0}
\renewcommand{\thefigure}{S\arabic{figure}}
\setcounter{table}{0}
\renewcommand{\thetable}{S\arabic{table}}

\section{Appendix A: Monotonic non-increasing hazard function for the log-logistic model}

We want to find which values of the log-logistic parameters give us monotonically non-increasing hazards. In a generic year, the hazard function is:
$$
h(a \mid \mu,\sigma) = \frac{\left(\frac{1}{\sigma\mu} \right) \left( \frac{a}{\mu}\right)^{\frac{1}{\sigma} - 1}}{1+\left(\frac{a}{\mu} \right)^{\frac{1}{\sigma}}}
$$
where $\mu>0$ and $\sigma>0$. To make the solution a bit more concise, use the shape parameter $\beta$, defined as $\beta = \sigma^{-1}$, so that the re-parameterized hazard is:
$$
h(a \mid \mu,\beta ) = \frac{\left(\frac{\beta}{\mu} \right) \left( \frac{a}{\mu}\right)^{\beta - 1}}{1+\left(\frac{a}{\mu} \right)^{\beta}}.
$$

To find when this function is monotonically non-increasing, we solve for the derivative and find when it is negative. Using the quotient rule and simplification, the derivative is:
\begin{align*}
    \frac{d}{da} h(a) &= \frac{\beta \left(\frac{1}{\mu}\right)^{\beta} a^{\beta-2}
\left[
\beta - \left(\frac{1}{\mu}\right)^{\beta} a^{\beta} -1
\right]}{D}
\end{align*}

where $D$ is a squared denominator and is hence always positive. To solve for when the derivative is negative, we can ignore $D$. We arrive at $\frac{d}{da}h(a) < 0$ when:
$$
\beta - \left(\frac{1}{\mu}\right)^\beta a^\beta < 1.
$$
Since $\mu > 0$ and $a\geq 0$, $\beta - \left(\frac{1}{\mu}\right)^\beta a^\beta \leq \beta$. So, if we choose $\beta<1$, we will necessarily get:
$$
\beta - \left(\frac{1}{\mu}\right)^\beta a^\beta \leq \beta < 1
$$
satisfying our requirement for a negative derivative. Therefore, a constraint that $\beta<1$ (or $\sigma>1$), in addition to the requirement that $\beta>0$, will give us a monotonically non-increasing hazard. To achieve this constraint, we fit $\theta_2$ where $\theta_2 = \text{logit}(1/\sigma)$.
\clearpage
\section{Appendix B: Missing mothers adjustment for HIV}

\subsection{Background and general approach}

For 17 countries where HIV prevalence exceeded 5 percent at any point in time since 1980, a missing mothers adjustment is applied by UN-IGME \citep{explanatory-igme,walker_child_2012}. This adjustment is applied to the direct estimates from DHS and MICS before B3 is run. To get the adjustment factors, a cohort component method of population projection (CCMP) like method is used. See \citet[Section 6.3]{preston2001demography} for more details on the CCMP method. Inputs come from Spectrum, and HIV positive and negative mothers and children are projected forward using assumptions such as with-HIV and HIV-free mortality rates for both mothers and children. Then U5MR is calculated from the simulated population using only children to mothers who are still alive (biased) and using all children (true). The ratio of these is the adjustment factor.

The objective of this appendix is to describe a method of adapting the missing mothers adjustment to the survival analysis framework we present. We begin by describing the general approach and then detail how the approach works for the piecewise-exponential and log-logistic survival models included in this paper. Finally, we illustrate the impact of these adjustments using the Kenya 2014 DHS.

We aim to use a parametric survival model for child mortality rates, in which a temporal model is applied to the survival model parameters as opposed to the mortality rates themselves. Therefore, we need to translate the adjustment factor for U5MR into a corresponding transformation of the survival model parameters. Suppose that our data come from the survival function $S^\star(a \mid \bm{\theta}^\star)$ defined by the parameters $\bm{\theta}^\star$, that the true survival function is $S(a\mid \bm{\theta})$ defined by the parameters $\bm{\theta}$, and that $S^\star(a \mid \bm{\theta}^\star)$ and $S(a \mid \bm{\theta})$ come from the same functional family. We start by using pseudo-likelihood estimation to produce $\widehat{\bm{\theta}}^\star$ and an estimate of its variance, $\widehat{\bm{V}}^\star$. Next we choose a transformation $f$ and assume that $\bm{\theta} = f(\bm{\theta}^\star)$. The exact transformation is different for each family of survival function but should generally achieve the result that U5MR is scaled by the same adjustment factor used by UN IGME. Finally, we use appropriate statistical methods for transformations to get $\widehat{\bm{\theta}}$ and the corresponding variance estimate $\widehat{\bm{V}}$, and use these adjusted values as input to the full Bayesian model instead of the unadjusted pseudo-likelihood estimates.

One significant difference between our approach and the current approach used by UN IGME is that no adjustment is currently applied to the NMR by UN IGME. This decision made by UN IGME is under the assumptions that neonatal mortality rate is not significantly associated with the HIV status of the mother \citep{fishel_child_2014} and HIV-specific mortality in neonates is negligible, and hence there should be little or no missing mother bias for NMR. We believe that it is reasonable to expect that children to mothers who have died from HIV would have higher all-cause mortality at all ages. However, for those applying the methods from this paper who prefer to not adjust NMR, a natural extension would be to use the piecewise-exponential model for countries where this bias is a big factor, and apply an adjustment where only the hazards after the first month are scaled.

\subsection{Piecewise-exponential}

Start with piecewise-exponential, which is relatively straightforward because the hazard has a linear form with respect to the model parameters. Suppose we have data that come from a piecewise-exponential with the following hazard function:
$$
h^\star(a \mid \bm{\theta} ) = \begin{cases}
\alpha_1^\star + \alpha_2^\star + \alpha_3^\star & \text{if } a \leq 1 \\
\alpha_1^\star + \alpha_2^\star & \text{if } 1<a\leq 12 \\
\alpha_1^\star & \text{if } 12<a\leq 60
\end{cases}
$$
where $\bm{\theta}^\star = \begin{bmatrix} \log(\alpha^\star_1) & \log(\alpha^\star_2) & \log(\alpha^\star_3) \end{bmatrix}^\top$.
when the truth which we want to make inference about has the hazard function:
$$
h(a \mid \bm{\theta}) = \begin{cases}
\alpha_1 + \alpha_2 + \alpha_3 & \text{if } a \leq 1 \\
\alpha_1 + \alpha_2 & \text{if } 1<a\leq 12 \\
\alpha_1 & \text{if } 12<a\leq 60,
\end{cases}
$$
where $\bm{\theta} = \begin{bmatrix} \log(\alpha_1) & \log(\alpha_2) & \log(\alpha_3) \end{bmatrix}^\top$.
We assume uniform multiplicative bias in the hazard function over age so that $h(a\mid \bm{\theta}) = b \times h^\star(a\mid \bm{\theta}^\star)$ for some bias parameter $b$. Now we can write $\alpha_1 = b\times \alpha_1^\star$, $\alpha_2 = b\times \alpha_2^\star$, and $\alpha_3 = b\cdot \alpha_3^\star$. Estimation of these parameters is done on the log-scale, so that we constrain the hazards to be positive.

From a single survey conducted in year $t'$ we obtain a vector of length $60 \times 1$ of pseudo-likelihood estimates:
$$\widehat{\bm{\theta}}^\star = \begin{bmatrix} \widehat{\bm{\theta}}^\star_{t'-19} & \cdots & \widehat{\bm{\theta}}^\star_{t'} \end{bmatrix}^\top,$$
where the survival parameters vary by year. We also obtain $\widehat{\bm{V}}^\star$, the $60 \times 60$ variance-covariance matrix for the pseudo-likelihood estimates for the survey. Each year has its own adjustment factor for U5MR in the original missing mother adjustment, and so we write $$\bm{b} = \begin{bmatrix} b_{t'-19} & b_{t'-19} & b_{t'-19} & \cdots & b_{t'} & b_{t'} & b_{t'} \end{bmatrix}^\top$$ as the $60 \times 1$ vector containing the set of proportionality constants for the transformations from $h^\star(a\mid \bm{\theta}^\star)$ to $h(a\mid \bm{\theta})$.

Since $\log(b\alpha) = \log(b) + \log(\alpha)$ we have $\bm{\theta}_t = \log(b_t) + \bm{\theta}^\star_t$ for an arbitrary year $t$. We assume that $\widehat{\bm{\theta}}^\star - \bm{\theta}^\star \sim \text{N}(\bm{0}, \widehat{\bm{V}}^\star)$. Adding and subtracting $\log(\bm{b})$ gives:
\begin{align*}
    \widehat{\bm{\theta}}^\star - \bm{\theta}^\star \mid  \widehat{\bm{V}}^\star &\sim \text{N}(\bm{0}, \widehat{\bm{V}}^\star) \\
    [\log(\bm{b})+\widehat{\bm{\theta}}^\star] - [\log(\bm{b})+\bm{\theta}^\star] \mid \widehat{\bm{V}}^\star &\sim \text{N}(\bm{0}, \widehat{\bm{V}}^\star) \\
    \widehat{\bm{\theta}} - \bm{\theta} \mid \widehat{\bm{V}}^\star &\sim \text{N}(\bm{0}, \widehat{\bm{V}}^\star).
\end{align*}
Hence, the transformed parameter vector has the same variance as the untransformed parameter vector, i.e.,~$\widehat{\bm{V}} = \widehat{\bm{V}}^\star$.

Given $\bm{b}$, we can complete this transformation and then proceed with $\widehat{\bm{\theta}}$ and $\widehat{\bm{V}}$ in our full temporal model as normal. Hence, it remains to determine what $\bm{b}$ is. We use adjustment factors provided by UN IGME, which are equivalent to:
\begin{equation}
   r_t = \frac{1-S_t(60 \mid \bm{\theta} )}{1-S_t^\star(60 \mid \bm{\theta}^\star)},
   \label{eq:adjustment_factor}
\end{equation}
the ratio of the true U5MR to the unadjusted U5MR, for $t = t'-19, \dots,t$. Via some algebra and the properties of survival models we obtain:
$$
\begin{aligned}
r_t &= \frac{1-S_t(60\mid \bm{\theta})}{1-S_t^\star(60 \mid \bm{\theta}^\star)} \\
&= \frac{1-\exp(-H_t(60\mid \bm{\theta}))}{1-\exp(-H_t^\star(60  \mid \bm{\theta}^\star ))} \\
&= \frac{1-\exp(-b_t H_t^\star(60  \mid \bm{\theta}^\star))}{1-\exp(-H_t^\star(60 \mid \bm{\theta}))} \\
\implies b_t &= -\frac{\log\left[1-r_t(1-\exp(-H_t^\star(60\mid \bm{\theta}^\star)) \right]}{H_t^\star(60\mid \bm{\theta}^\star)} \\
&= \frac{\log\left[1-r_t(1-S_t^\star(60\mid \bm{\theta^\star})) \right]}{\log[S_t^\star(60\mid \bm{\theta}^\star)]} ,
\end{aligned}
$$
where $H_t^\star(60  \mid \bm{\theta}^\star )= \int_0^x h_t^\star(x  \mid \bm{\theta}^\star) ~dx$ and $H_t^\star(60  \mid \bm{\theta} )= \int_0^x h_t(x  \mid \bm{\theta}) ~dx$.
Since we have $r_t$ from UN IGME and $S_t^\star(60)$ from the pseudo-likelihood estimates, we can use this equation to obtain $b_t$, for $t =t'-19,\dots,t'$.

\subsection{Log-logistic model}

From an arbitrary survey conducted in year $t'$, we get the $40 \times 1$ vector of log-logistic pseudo-likelihood estimates $\widehat{\bm{\theta}}^\star = \begin{bmatrix} \widehat{\bm{\theta}}^\star_{t'-19} & \dots & \widehat{\bm{\theta}}^\star_{t'} \end{bmatrix}^\top$, where for $t =t'-19,\dots,t'$:
$$
\widehat{\bm{\theta}}_t^\star =\begin{bmatrix}
    \widehat\theta_{t1}^\star\\ \widehat\theta_{t2}^\star
\end{bmatrix} =  \begin{bmatrix}
    \log(\widehat{\mu}_t^\star) \\ \text{logit}(1/\widehat{\sigma}_t^\star)
\end{bmatrix}.
$$
To transform these observed parameters to arrive at log-logistic parameters on the scale we want to make inference on, we assume:
$$
\widehat{\bm{\theta}}_t = \begin{bmatrix}
    \widehat\theta_{t1}\\\widehat \theta_{t2}
\end{bmatrix} =  \begin{bmatrix}
    \log(b_t \widehat{\mu}_t^\star) \\ \text{logit}(1/\widehat{\sigma}_t^\star)
\end{bmatrix}
$$
so that $\sigma_t = \sigma_t^\star$ is fixed but $\mu_t = b_t \mu_t^\star$ for some scalar $b_t$. We selected this particular approach because it is a simple transformation that leads to an approximately equal hazard ratio with respect to age when comparing the adjusted and unadjusted hazard functions. We illustrate this with an example in the next section. We assume the UN IGME adjustment factor to U5MR, $r_t$, for each year $t =t'-19,\dots,t'$, applies. The adjustment factor is defined as in Equation \eqref{eq:adjustment_factor}. Since the log-logistic survival function has the form:
$$S_t(60 \mid \bm{\theta})=\frac{1}{1+\left(60/\mu_t\right)^{1/\sigma_t}}$$
which means $S_t(60 \mid \bm{\theta}) = S_t^\star(60/b_t \mid \bm{\theta}^\star)$. Hence,
$$
r_t = \frac{1-S_t^\star(60/b_t\mid \bm{\theta}^\star)}{1-S_t^\star(60\mid \bm{\theta})}.
$$
After we obtain $\widehat \mu_t^\star$ and $\widehat \sigma_t^\star$, we use this equation plus some algebra to find $b_t$. This results in:
$$
b_t = \frac{60}{(S_t^\star)^{-1}(1-r_t[1-S_t^\star(60)])}
$$
where the inverse survival function is:
$$
(S_t^\star)^{-1}(s) = \mu_t^\star\left(\frac{1}{s}-1\right)^{\sigma_t^\star}.
$$
Using the same logic as for the piecewise-exponential, we find that $\widehat{\bm{V}} = \widehat{\bm{V}}^\star$ for the log-logistic as well, when we apply the transformation we've defined.

\subsection{Illustration for Kenya 2014 DHS}

\begin{figure}
    \centering
    \includegraphics[width=0.9\linewidth]{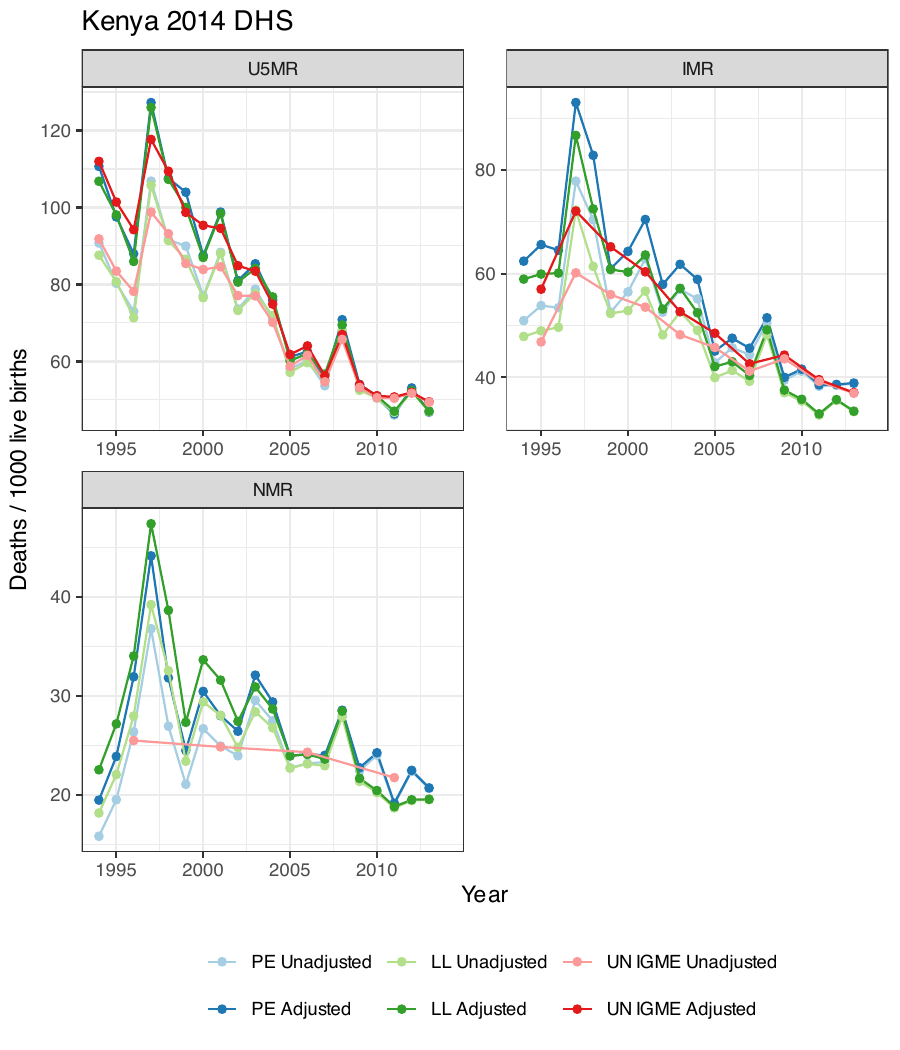}
    \caption{Under-five, infant, and neonatal mortality rates in Kenya based on the 2014 DHS, pre- and post-missing-mothers adjustment. PE = Piecewise-exponential; LL = log-logistic.}
    \label{fig:mm_adjustment}
\end{figure}

We illustrate the impact of the missing mothers transformation on the hazard function and on the mortality rates of interest using the observations from the Kenya 2014 DHS. Using pseudo-likelihood estimation in combination with the adjustment methods described in this appendix, we get the results in Table \ref{tab:mm_adjustment} (showing the year 2000 only for illustration) and Figure \ref{fig:mm_adjustment}, with the same data as processed by UN IGME included for comparison. However, note that no adjustment is made for NMR by UN IGME and they do not incorporate a direct weighted estimate of IMR or NMR for single years such as 2000 from DHS, so there are no entries in Table \ref{tab:mm_adjustment} for UN IGME for those age groups.

\begin{table}[ht]
\centering
\begin{tabular}{llccc}
\hline
& Indicator & Unadjusted & Adjusted & Ratio \\
\hline
UN IGME & U5MR & 83.9 & 95.4 & 1.137\\
\hline
Piecewise-exponential & U5MR & 77.0 & 87.5 & 1.137 \\
& IMR & 56.5 & 64.3 & 1.138 \\
& NMR & 26.7 & 30.5 & 1.141 \\
\hline
Log-logistic & U5MR & 76.6 & 87.0 & 1.137 \\
& IMR  & 52.9 & 60.3 & 1.141 \\
& NMR  & 29.4 & 33.7 & 1.145 \\
\hline
\end{tabular}
\caption{Results for missing mothers adjustment for estimates of mortality rates in Kenya in the year 2000 based on the 2014 DHS.}
\label{tab:mm_adjustment}
\end{table}
\clearpage
\section{Appendix C: Estimating the standard error associated with the Brass summary birth history method for child mortality}

\subsection{Background and objective}

One data type which is utilized by UN IGME to estimate child mortality rates is summary birth histories (SBH). SBHs are inexpensive and quick to collect as compared to full birth histories, and may come from censuses or household surveys. A SBH includes counts of children ever born to women surveyed, as well as the number of those children who have died. Analysts apply demographic methods to estimate child mortality rates from these summaries. The classic approach is the Brass method based on the mother's age at the time of the survey \citep{brass1968methods,trussell1975brass}. Improvements based on the time since a mother's first birth have been proposed, but time since first birth is not typically available in older censuses. Other SBH methods exist, such as those described by \citet{wilson2022probabilistic}, \citet{rajaratnam2010measuring}, and \citet{burstein2018newmethod} but these methods are not used by UN IGME and are not the focus of this analysis.

To incorporate a SBH estimate of child mortality into our larger Bayesian model for child mortality rates over time we use the likelihood:
\begin{equation}
\text{logit}({_a\widehat{q}_{0,t}}) \sim \text{N}(\text{logit}({_aq_{0,t}}), \sigma_{a,t}^2)
\end{equation}
where ${_a\widehat{q}_{0,t}}$ is a Brass estimate of the probability of death by age $a$ in year $t$ and ${_aq_{0,t}}$ is the true probability of death by age $a$ in year $t$. The choice of $\sigma_{a,t}^2$, the variance of this estimate, is not straightforward, because the Brass method is deterministic and not a statistical model. The purpose of the analysis in this appendix is to present a rigorous selection of $\sigma_{a,t}^2$ for use within our child mortality model. The current procedure used by UN IGME is to apply a coefficient of variation of 0.025 for census-based SBH estimates and 0.1 for household-survey based SBH estimates. There is no documentation that we are aware of for how these values were chosen, but they originated with the B3 paper by \citet{alkema2014global}.

First consider the relatively simple case where the SBH comes from a complete census. We can think of $\sigma_{a,t}^2$ as the measurement error associated with the Brass method. Let $\mathcal{A}$ be the complete set of survival information, including dates of birth and dates of death, for children born during a time period of interest. If we had access to $\mathcal{A}$, such as from vital registration, we could directly compute observed mortality rates. However, for SBH, we don't have $\mathcal{A}$, but rather the data summary $T(\mathcal{A})$, which is the number of children ever born and the number of those who have died, by age group of the mother. Many realities of $\mathcal{A}$ can lead to the same $T(\mathcal{A})$. The Brass method aims to identify the past mortality rates most likely to give rise to $T(\mathcal{A})$, however, because there are many possibilities for the truth given SBH data, the Brass method is an imperfect measurement tool with some precision we would like to estimate. For household survey-based SBHs, Brass estimates are subject to additional sampling error. This sampling error can be obtained using jackknife \citep{wilson2022probabilistic,pedersen2012timeperiods}.

\subsection{Methods}

\subsubsection{Generating SBH data}

\begin{figure*}[!h]
    \centering
    \includegraphics[width=1\linewidth]{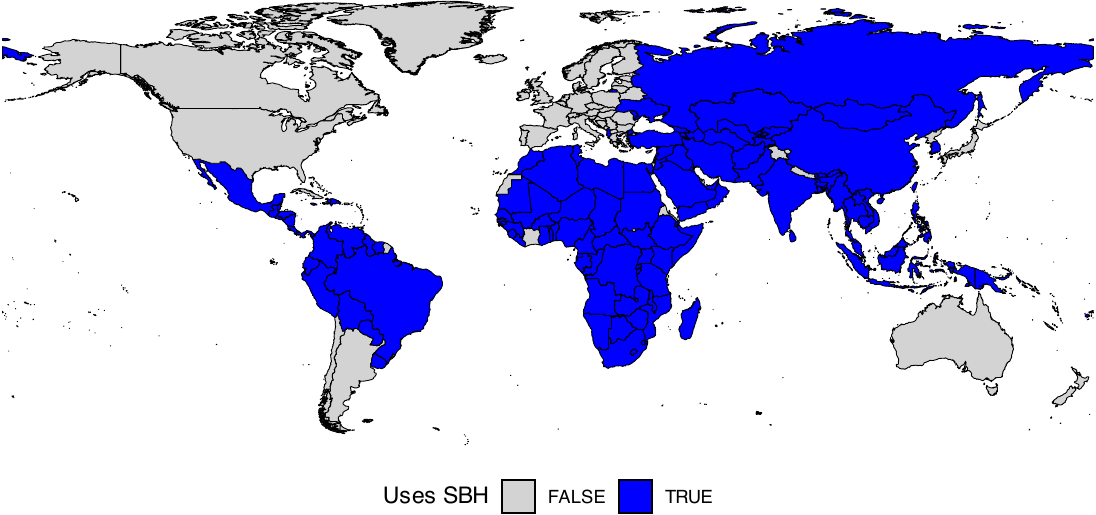}
    \caption{Countries with summary birth history (SBH) data included in this analysis.}
    \label{fig:sbh-map}
\end{figure*}

Annual estimates of births, fraction of births by mother's age, mortality rates by single year of age, and population, from the World Population Prospects 2024 study were downloaded using the {\tt wpp2024} {\tt R} package (https://github.com/PPgp/wpp2024) \citep{unpd2024wpp}. The years included were 1950--2019 and the countries included were the 133 countries with SBH appearing  in the UN IGME 2023 estimates of child mortality (Figure \ref{fig:sbh-map}). The WPP estimates are taken to represent true populations for which we would like to make estimates using the Brass method. We generate synthetic SBH data for these populations as described in this section, then estimate IMR and U5MR, and compare those estimates to the mortality rates that gave rise to the SBH data. Since the WPP values are themselves estimates of true populations and not real data, we recognize that the populations we are studying may differ somewhat from the real life populations they are based on. However, this should not affect this analysis because the SBH data in this study are not real observations but rather are derived from the WPP estimates we are comparing them to.

Using the WPP estimates, we generated SBH data for hypothetical censuses in 2000, 2010, and 2019. No census year prior to 2000 was used, because we need information about womens' survival and fertility starting with their birth and ending when they reach age 50. This is 50 years in total and the WPP estimates begin at 1950. No census years beyond 2019 were considered because of the COVID-19 pandemic. SBH methods are known to perform poorly in the presence of shocks, and do best when there is a gradual and steady temporal trend in mortality \citep{silva2012consistency}. We chose three census years approximately 10 years apart rather than every year between 2000 and 2019 because results from adjacent years are likely to be highly correlated and not provide much additional information.

For a hypothetical census, let $m$ denote a maternal age at the time of the census in years, and $t$ denote the year of the census. Then the number of children born to the birth cohort of women aged $m$ in year $t$, by the end of year $t$, is
\begin{equation}
\text{CEB}_{mt} = \sum_{a=0}^m B_{t-m+a} \times c_{t-m+a,a}
\label{eq:sbh2}
\end{equation}
where $B_y$ is the number of births in year $y$ and $c_{y,z}$ is the proportion of births in year $y$ to women aged $z$. Next, the number of these children who have died before the end of year $t$ is:
\begin{equation}
\text{CD}_{mt} = \sum_{a=0}^m B_{t-m+a} \times c_{t-m+a,a} \times \left\{ 1-\prod_{k=0}^{m-a} \left(1 - {_{t-m+a+k}q_{0,k}}\right) \right\}.
\label{eq:sbh3}
\end{equation}
In a real census, births and deaths for children to women who died before the time of the census would not be counted. To address this, we assume mortality risk for mothers is not associated with either their fertility or the mortality risk of their children. Then, the recorded children ever born and children died is equal to the total from equations \eqref{eq:sbh2} or \eqref{eq:sbh3} times the probability a woman born in year $t-m$ would survive to age $m$. The other piece of information required for the Brass method is the number of women by age at the time of the census -- this comes directly from WPP population estimates. Finally, the Brass method uses the set of aggregated age groups $\mathcal{A} = \{\text{15--19, 20--24, 25--29, 30--34, 35--39, 40--44, 45--49}\}$. To get counts of children ever born, children died, and women by age group in $\mathcal{A}$ we simply sum over the single-year age groups.

\subsubsection{Brass summary birth history estimates}

For each synthetic census, the Brass estimates are obtained based on tables compiled by \citet{trussell1975brass}. The method relies on a selection of a model life table family among the Coale Demeny "north", "south", "east", and "west" regions. We selected the model life table region by finding the life table with the closest ${_5q_0}$ to the true ${_5q_0}$ within each region, and then selecting the region with the closest ${_1q_0}/{_4q_1}$ ratio to the true ratio. For each census year $t \in \{2000, 2010, 2019\}$ and mothers age group $i \in \mathcal{A}$, the Brass method produces one estimate of IMR and one estimate of U5MR for some reference time period before year $t$. Estimates generated from younger age groups are dated to earlier reference periods, closer to the time of the census.

\subsubsection{Computing error}

From the collection of synthetic censuses, let $\widehat{q}_{x,i,j}$ be the $j$-th SBH estimate of the probability of death by age $x$ based on mothers in age group $i$ for $x \in \{1,5\}$, $i \in \mathcal{A}$, and $j = 1, \ldots, J_{x,i}$. Suppose $q_{x,i,j}$ is the true mortality probability from WPP for age $x$ and for the same year as the reference year assigned to $\widehat{q}_{x,i,j}$ by the Brass method.

Define the residual for an observation as
\begin{equation}
r_{x,i,j} = \text{logit}(\widehat{q}_{x,i,j}) - \text{logit}(q_{x,i,j})
\end{equation}
and let the mean residual for child's age $x$ and mothers age group $i$ be $\bar{r}_{x,i} = \frac{1}{J_{x,i}} \sum_{j=1}^{J_{x,i}} r_{x,i,j}$. Compute a standard error for age $x$ and mother's age $i$ as the standard deviation of these residuals:
\begin{equation}
\sigma_{x,i} = \sqrt{\frac{1}{J_{x,i}-1}\sum_{i=1}^{J_{x,i}}(r_{x,i,j} - \bar{r}_{x,i})^2}.
\end{equation}
Unlike the current practice by UN IGME, we choose to stratify this by age of the child and mothers age group because we observe notable differences across these parameters. To validate this choice of standard error, we test the coverage of the resulting 90\% intervals.

By using the delta method, we can estimate the standard error in regular-space from the standard error in logit-space. Then we can divide this standard error by each SBH estimate to get the coefficient of variation of each observation. The purpose of this step is to facilitate comparison with the UN IGME approach of using a coefficient of variation of 0.025 for census-based SBH.

\subsection{Results}

\begin{figure*}[!h]
    \centering
    \includegraphics[scale=0.65]{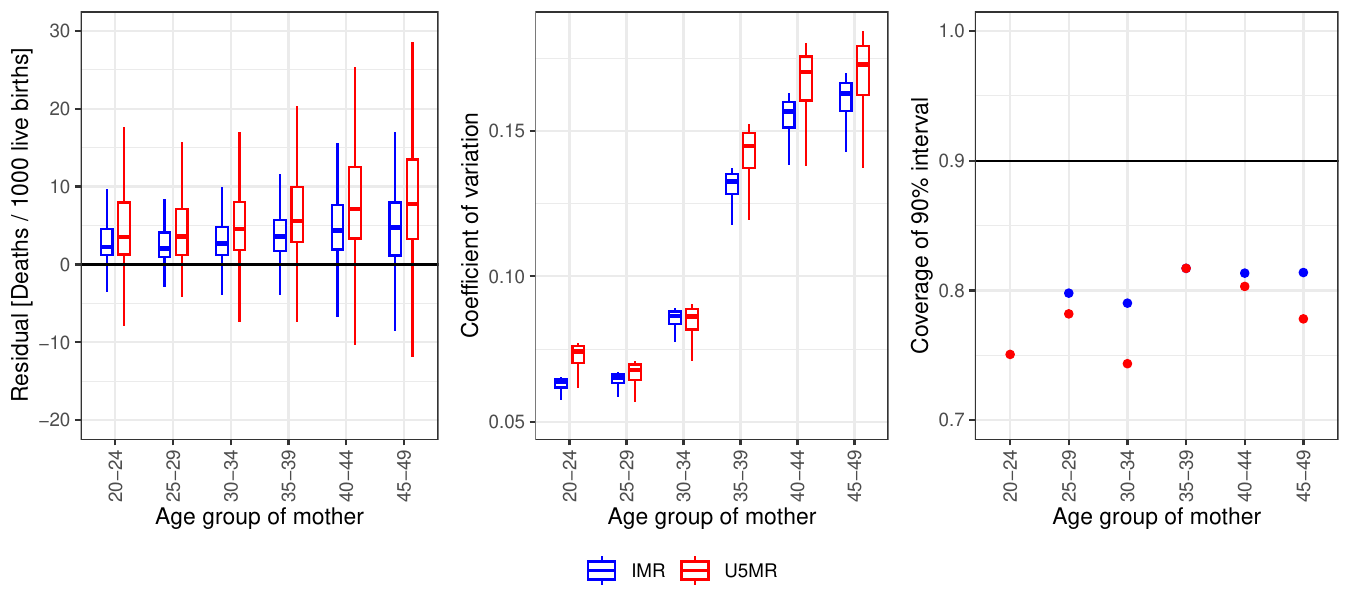}
    \caption{Residuals in units of deaths per 1000 live births, coefficient of variation, and coverage of 90\% intervals, each by the age group of the mother at the time of the survey and for infant mortality rate (IMR) and under-five mortality rate (U5MR) separately.}
    \label{fig:sbh-results}
\end{figure*}

The distribution of residuals is presented in Figure \ref{fig:sbh-results}. Generally speaking, we find that the Brass method over-estimates mortality rates, but typically by fewer than 10 deaths per 1000 live births. This finding is consistent with the results reported by \cite{verhulst_child_2016}. Per \citet{verhulst_child_2016} and \citet{hill2013direct}, a global declining pattern in fertility is responsible. The parity ratios which are inputs to the Brass method will be inaccurate in an environment of changing fertility rates, because they are based on the time of the census. See \citet{verhulst_child_2016} for an extended explanation for how this leads to over-estimation. Note that results for mothers aged 15--19 are not used by UN IGME because they are considered unreliable due to the relatively high mortality of children born to the youngest mothers.

The coefficient of variation we estimate is uniformly above the UN IGME level of 0.025 (and often considerably greater), and we also find that it has a strong association with the mothers age (Figure \ref{fig:sbh-results}). The coefficient of variation for age 20--24, 25--29, and 30--34 is between 5\% and 10\%, whereas the coefficient of variation is closer to 12\% for 35-39 and between 15\% and 20\% for 40-44 and 45-49. This is consistent with the consensus that estimates from age groups 20-24, 25-29, and 30-34 are the most reliable. There are some differences in coefficient of variation between IMR and U5MR but these differences are small compared to the differences across age group of the mother.

Coverage of the 90\% intervals computed using our methods is below the nominal 90\%, but above 70\% for each group (Figure \ref{fig:sbh-results}). Lower-than-nominal coverage might be expected given the degree of bias present, even if we have accurately captured the variance of the estimates. These findings support our decision to use a higher standard error in our Bayesian model than what is currently being used by UN IGME. The final variance which is used for the likelihood for census-based SBH observations in our full Bayesian model is presented in Table \ref{tab:sbh}. For survey-based SBH we use a 10\% coefficient of variation based on the current practice by UN IGME but we suggest future applications of these methods re-calculate SBH standard error using the jackknife \citep{pedersen2012timeperiods}, and possibly adding the sampling variance as estimated by the jackknife to additional variance coming from the measurement precision as described in this appendix.

\begin{table*}[!h]
\centering
\begin{tabular}[t]{lrrr}
\toprule
Age of mother & SE logit(IMR) & SE logit(U5MR) & Covariance \\
\midrule
20--24 & 0.066 & 0.078 & 0.0039 \\
25--29 & 0.068 & 0.072 & 0.0035\\
30--34 & 0.090 & 0.091 & 0.0066\\
35--39 & 0.139 & 0.154 & 0.0191\\
40--44 & 0.164 & 0.182 & 0.0272\\
45--49 & 0.172 & 0.186 & 0.0290\\
\bottomrule
\end{tabular}
\caption{Standard error and covariance for logit(IMR) and logit(U5MR), by age of the mother.}
\label{tab:sbh}
\end{table*}

\clearpage

\section{Appendix D: Supplemental Results}

This supplement contains a collection of additional figures, including a summary of the estimated survival parameters across surveys (from pseudo-likelihood estimation) and vital registration (from maximum-likelihood estimation), as well as additional results for Kenya, Brazil, Estonia, and Syrian Arab Republic.

\begin{figure}[ht]
    \centering
    \begin{subfigure}{0.8\textwidth}
        \centering
        \includegraphics[width=\linewidth]{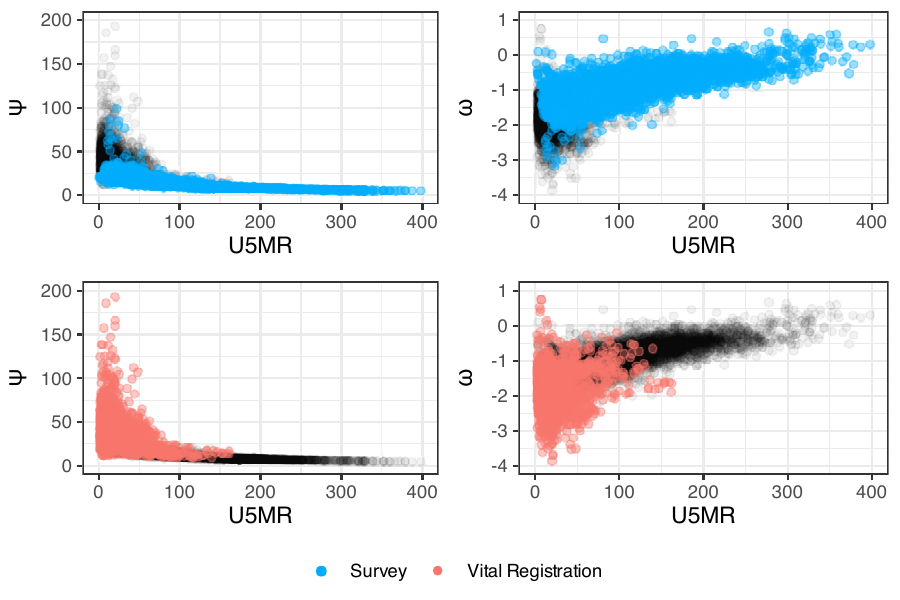}
        \caption{Log-logistic}
        \label{fig:first}
    \vspace{1em}
    \end{subfigure}
    
    \begin{subfigure}{0.8\textwidth}
        \centering
        \includegraphics[width=\linewidth]{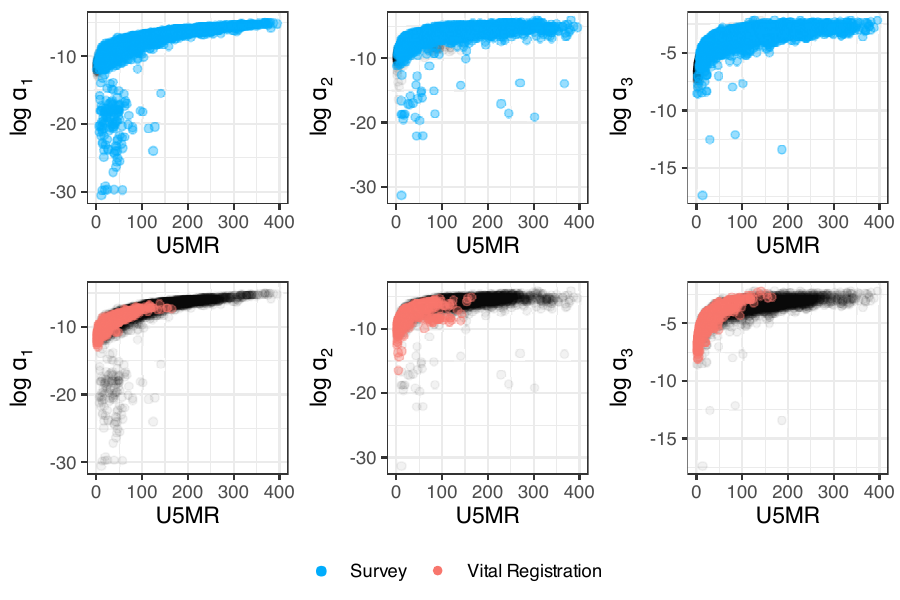}
        \caption{Piecewise-exponential}
        \label{fig:second}
    \end{subfigure}
    \caption{Fitted estimates of the log-logistic parameters and the piecewise-exponential parameters, using pseudo-likelihood estimation for survey data from DHS and MICS and maximum likelihood estimation for count data from vital registration. Survey or vital registration results are plotted as faded black points for panels where they are not the focus.}
    \label{fig:par-ests-all}
\end{figure}

\begin{figure}
    \centering
    \includegraphics[width=0.9\linewidth]{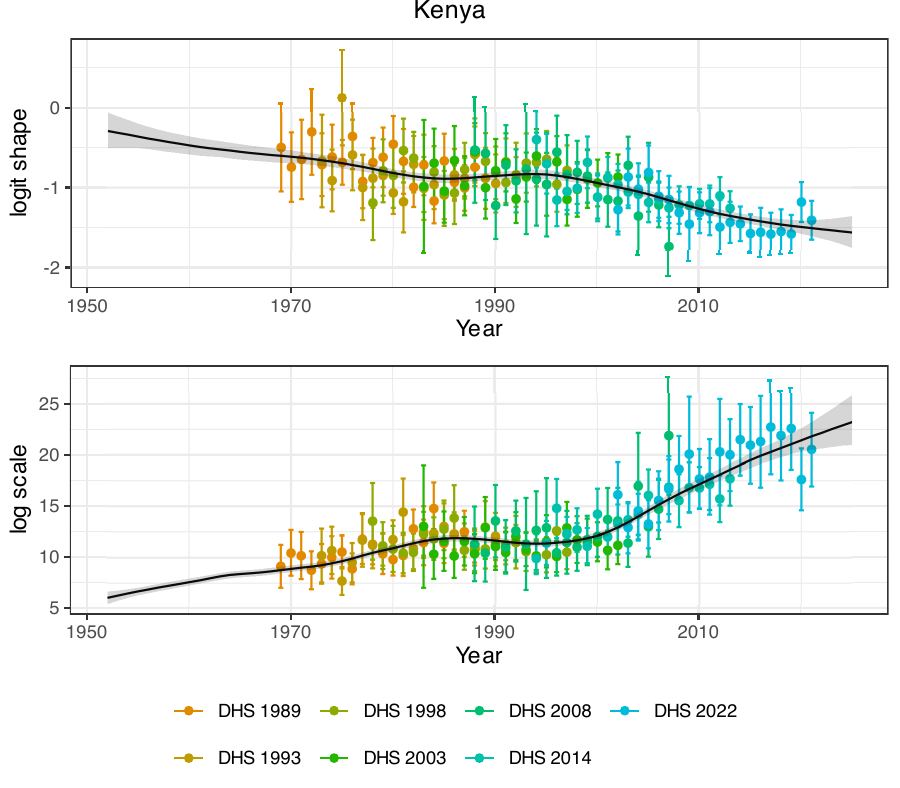}
    \caption{Estimates of survival parameters for the log-logistic model, for Kenya.}
    \label{fig:kenya-pars-ll}
\end{figure}

\begin{figure}
    \centering
    \includegraphics[width=0.9\linewidth]{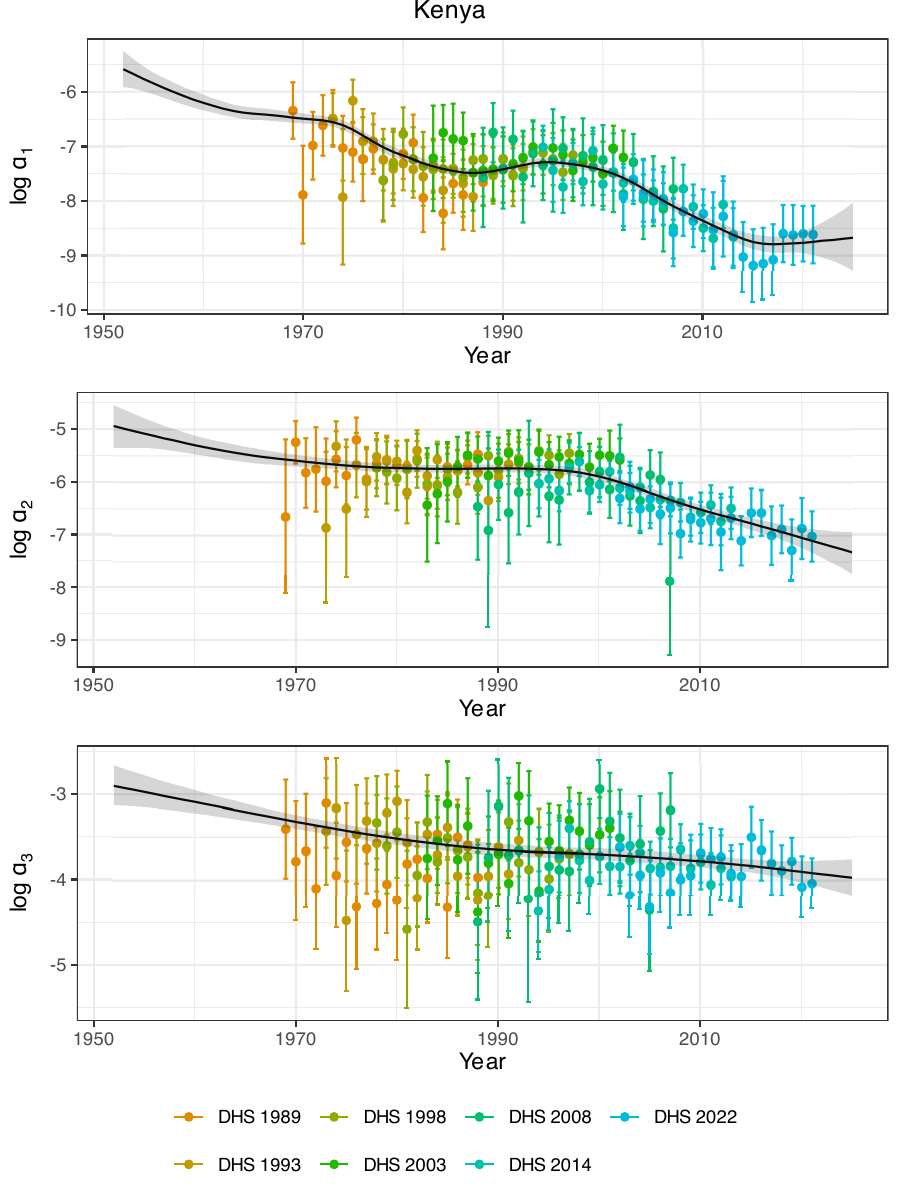}
    \caption{Estimates of survival parameters, for the piecewise-exponential model, for Kenya.}
    \label{fig:kenya-pars-pe}
\end{figure}

\begin{figure}
    \centering
    \includegraphics[width=0.9\linewidth]{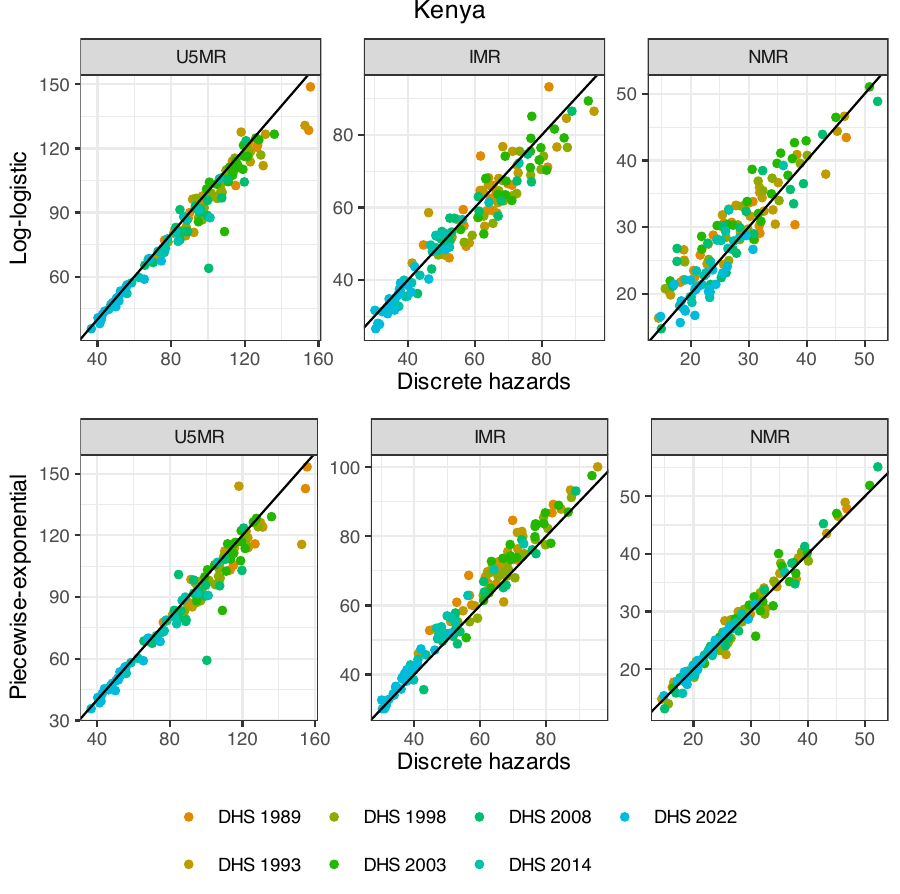}
    \caption{Comparison of discrete hazards estimates to pseudo-likelihood estimates from surveys, for Kenya.}
    \label{fig:kenya-comp-to-dh}
\end{figure}

\begin{figure}
    \centering
    \includegraphics[width=0.9\linewidth]{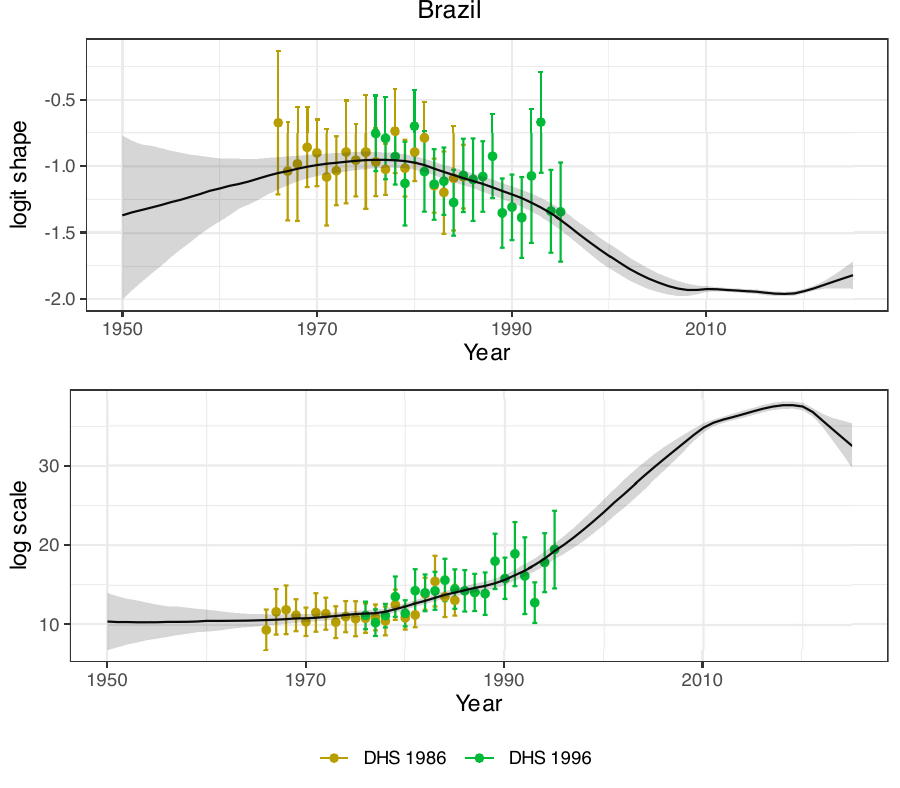}
    \caption{Estimates of survival parameters for the log-logistic model, for Brazil.}
    \label{fig:brazil-pars-ll}
\end{figure}

\begin{figure}
    \centering
    \includegraphics[width=0.9\linewidth]{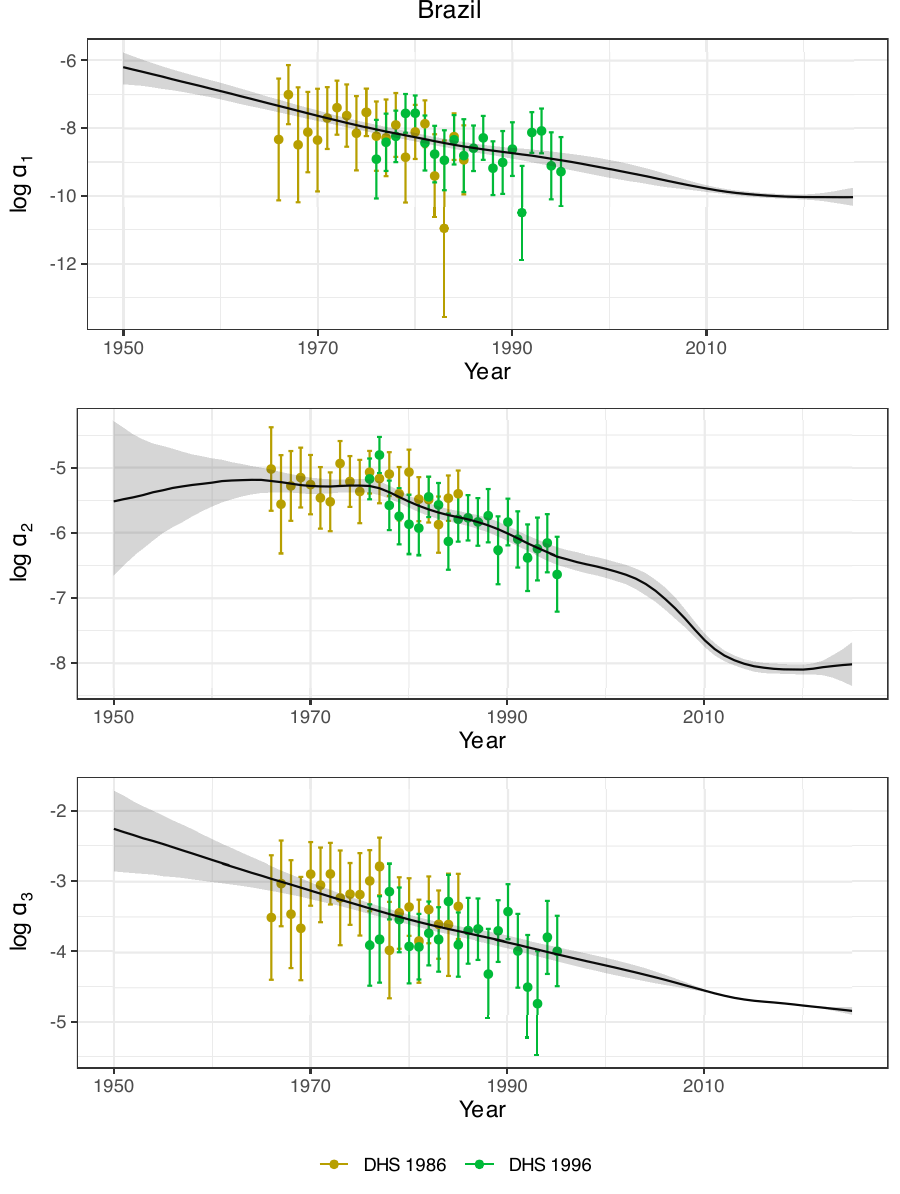}
    \caption{Estimates of survival parameters for the piecewise-exponential model, for Brazil.}
    \label{fig:brazil-pars-pe}
\end{figure}

\begin{figure}
    \centering
    \includegraphics[width=0.9\linewidth]{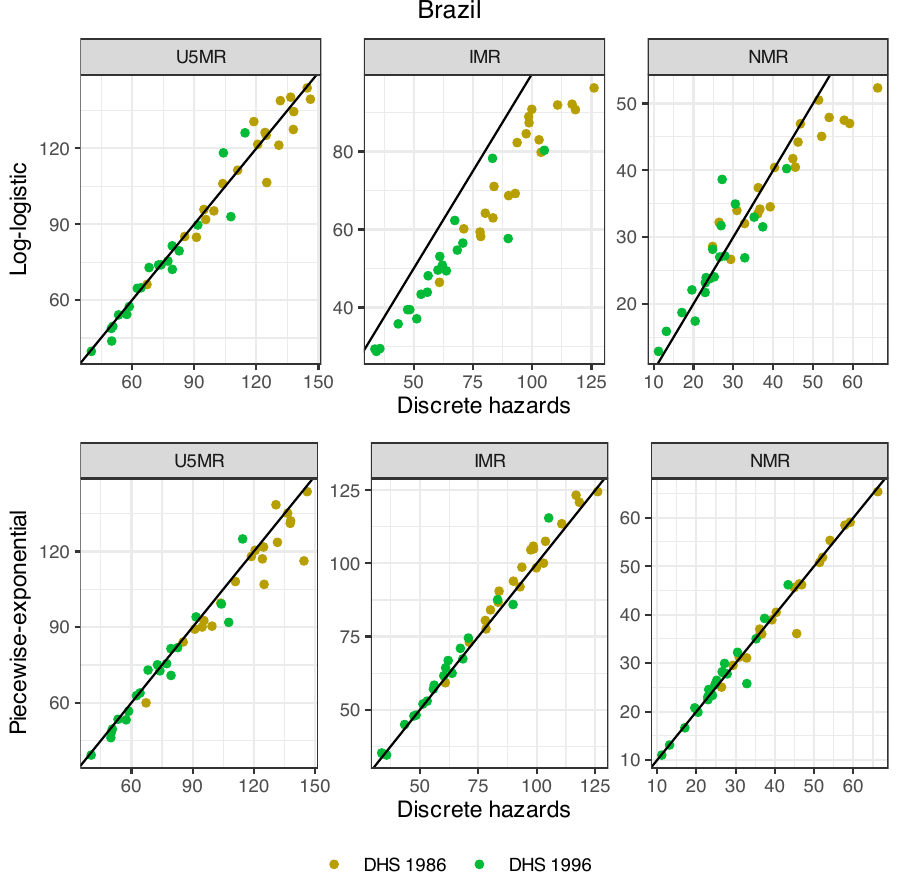}
    \caption{Comparison of discrete hazards estimates to pseudo-likelihood estimates from surveys, for Brazil.}
    \label{fig:brazil-comp-to-dh}
\end{figure}

\begin{figure}
    \centering
    \includegraphics[width=0.9\linewidth]{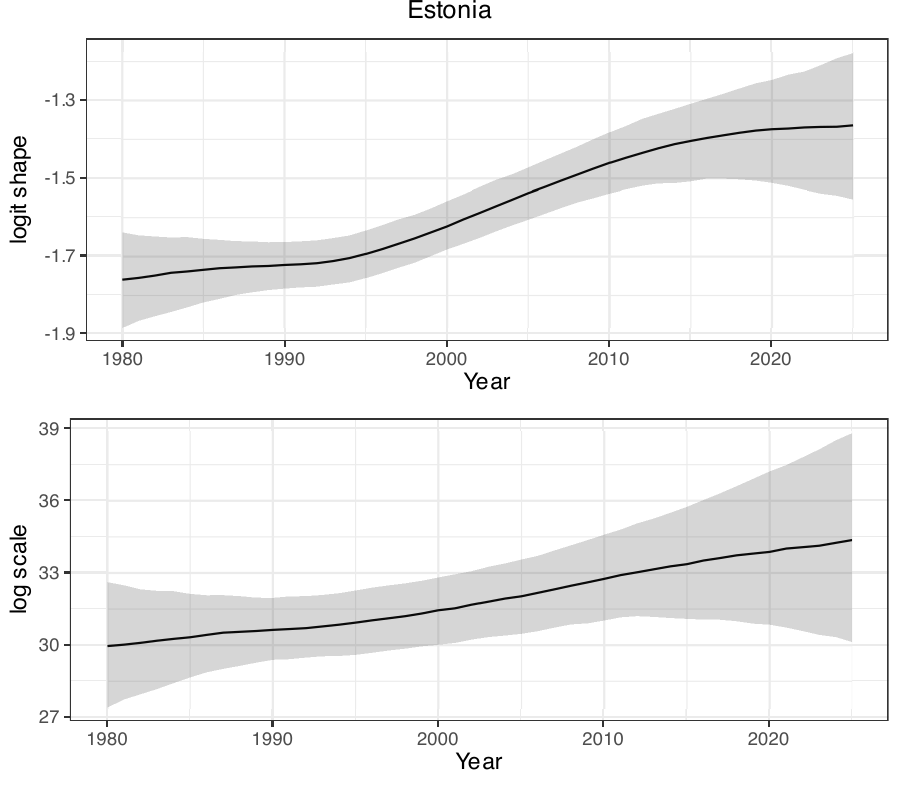}
    \caption{Estimates of survival parameters for the log-logistic model, for Estonia.}
    \label{fig:estonia-pars-ll}
\end{figure}

\begin{figure}
    \centering
    \includegraphics[width=0.9\linewidth]{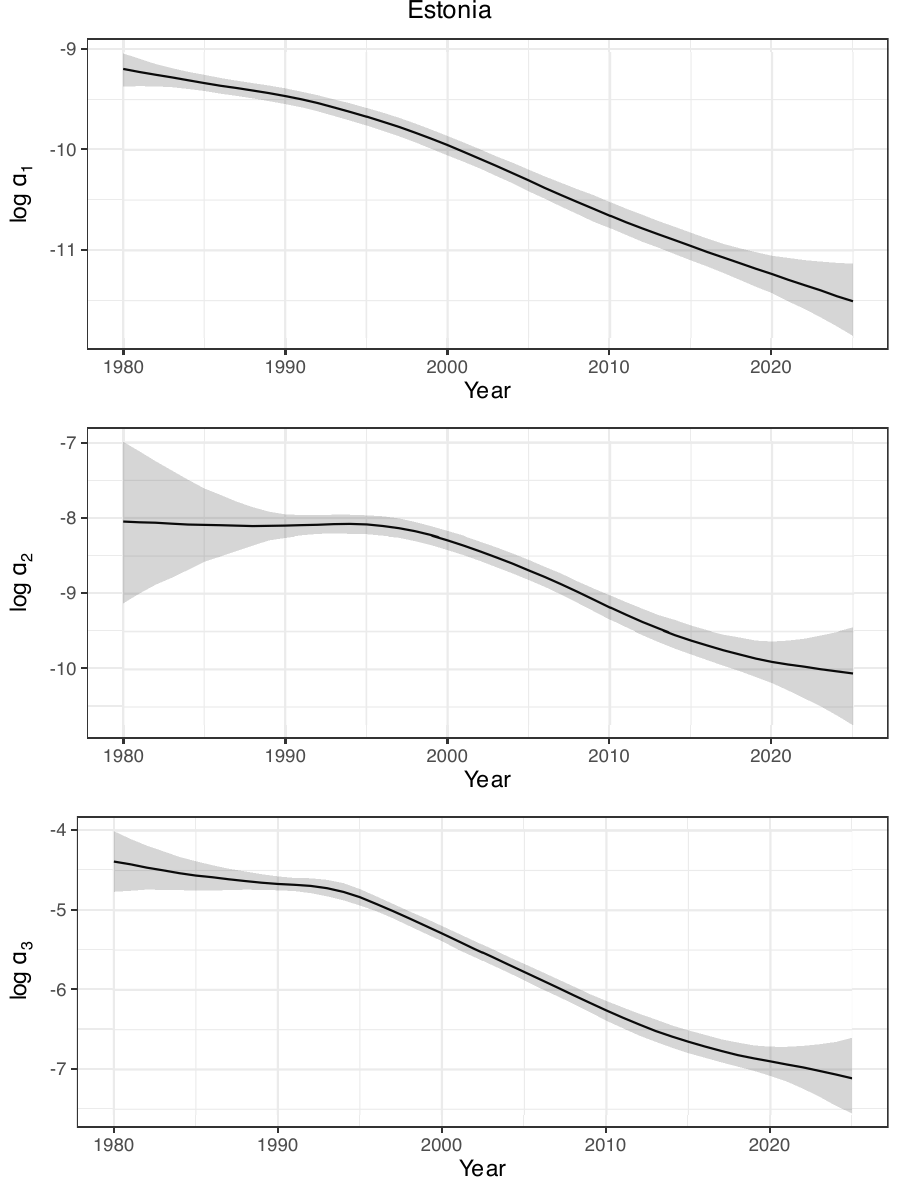}
    \caption{Estimates of survival parameters for the piecewise-exponential model, for Estonia.}
    \label{fig:estonia-pars-pe}
\end{figure}

\begin{figure}
    \centering
    \includegraphics[width=0.9\linewidth]{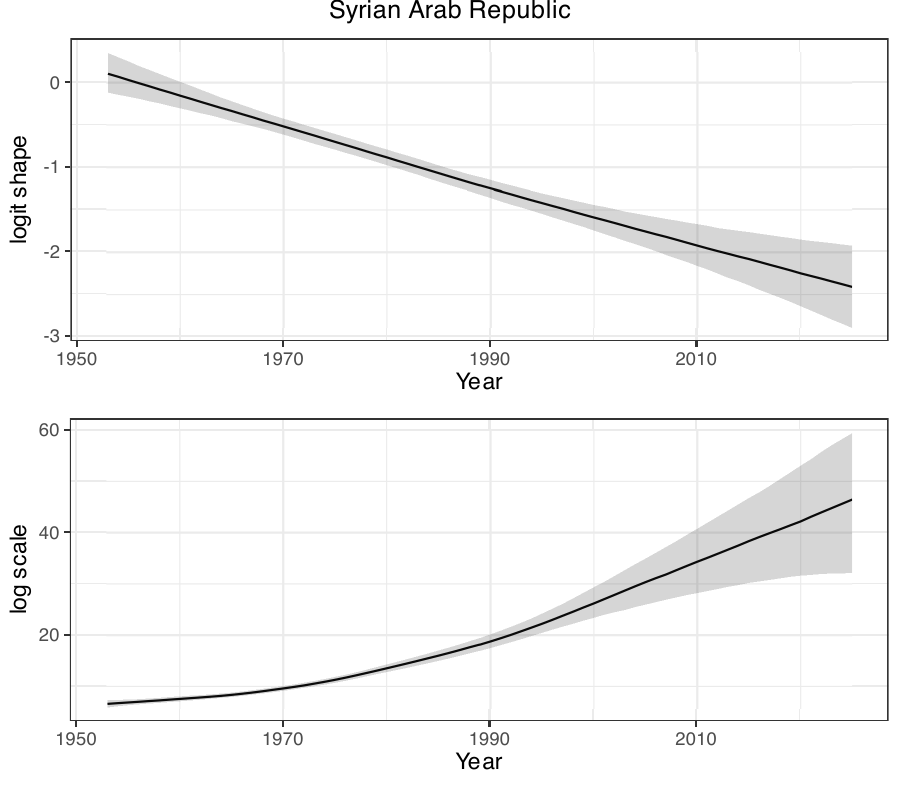}
    \caption{Estimates of survival parameters for the log-logistic model, for Syrian Arab Republic.}
    \label{fig:syria-pars-ll}
\end{figure}

\begin{figure}
    \centering
    \includegraphics[width=0.9\linewidth]{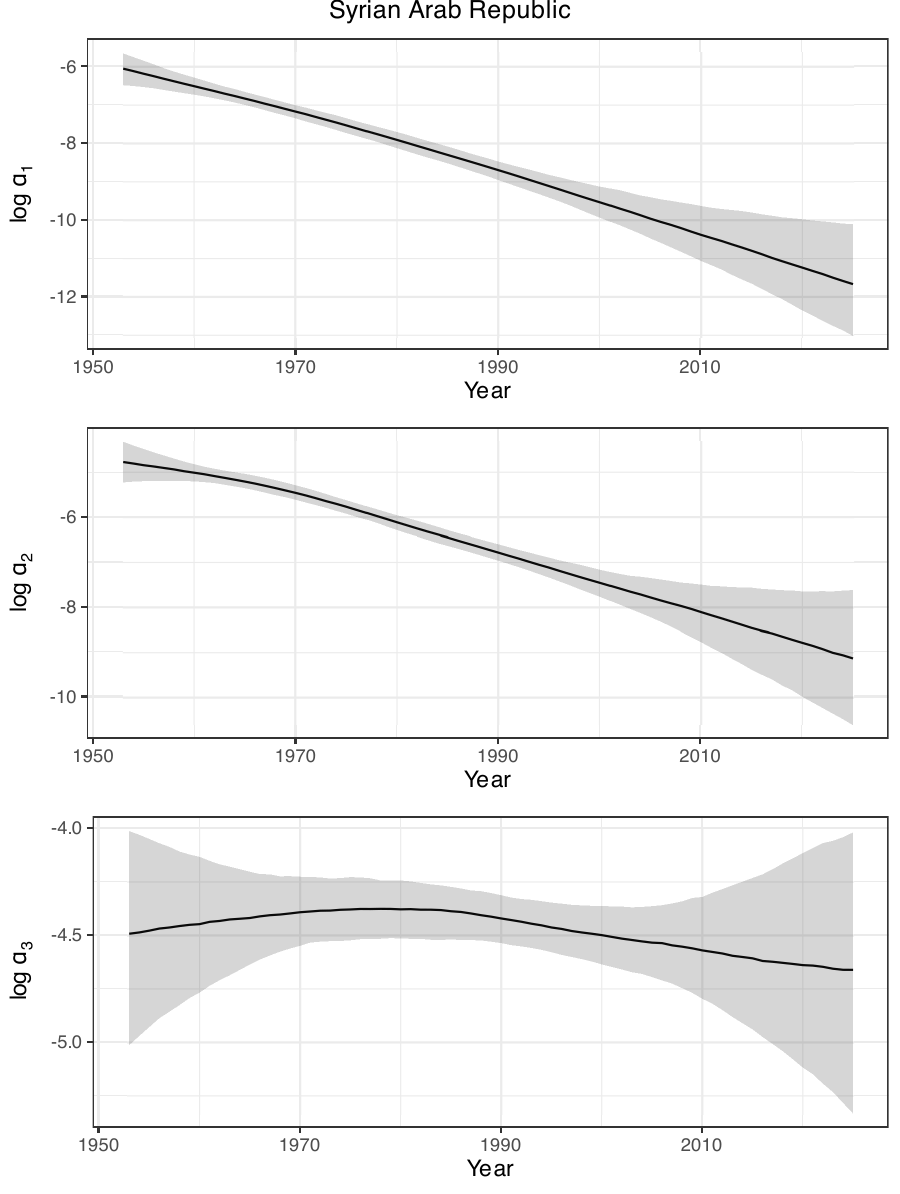}
    \caption{Estimates of survival parameters for the piecewise-exponential model, for Syrian Arab Republic.}
    \label{fig:syria-pars-pe}
\end{figure}

\begin{figure}
    \centering
    \includegraphics[width=0.9\linewidth]{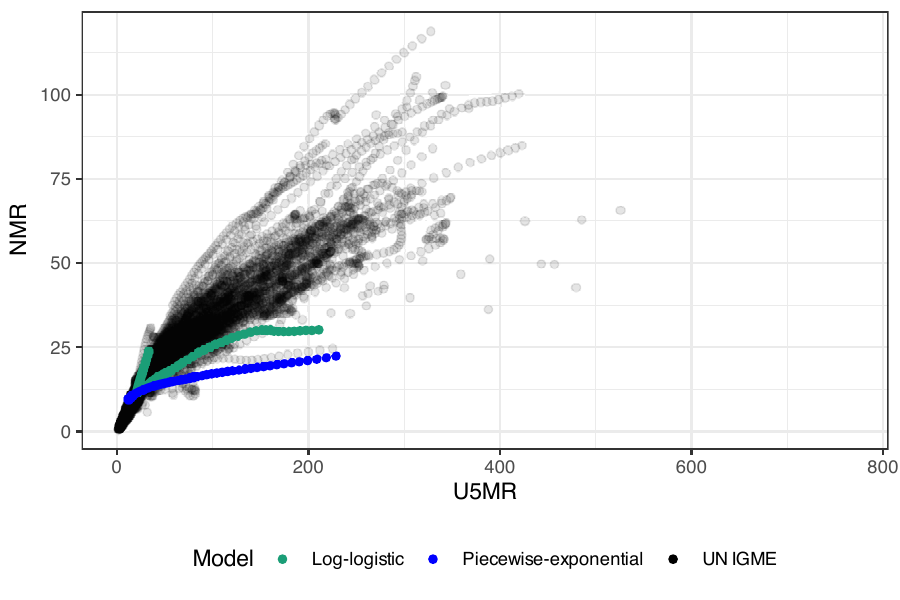}
    \caption{Comparison of NMR to U5MR coming from UN IGME estimates, and coming from either the log-logistic or the piecewise-exponential model for Syrian Arab Republic.}
    \label{fig:syria-nmr-u5mr-scatter}
\end{figure}

\end{document}